\documentclass[]{aastex631}

\usepackage{bm}
\usepackage{amsmath}

\newcommand\ltsima{$\; \buildrel <\over\sim \;$}
\newcommand\simlt{\lower.5ex\hbox{\ltsima}}
\newcommand\gtsima{$\; \buildrel >\over\sim \;$}
\newcommand\simgt{\lower.5ex\hbox{\gtsima}}

\shorttitle{Event Rate and Optical Depth from MOA-II 9 yr}
\shortauthors{Nunota et al.}

\graphicspath{{./}{figures/}}

\defcitealias{suz16}{S16}
\defcitealias{kos23}{K23}

\begin{document}

\title{The Microlensing Event Rate and Optical Depth from MOA-II 9 year Survey toward the Galactic Bulge}

\author{Kansuke Nunota}
\affiliation{Department of Earth and Space Science, Graduate School of Science, Osaka University, Toyonaka, Osaka 560-0043, Japan}

\author{Takahiro Sumi}
\affiliation{Department of Earth and Space Science, Graduate School of Science, Osaka University, Toyonaka, Osaka 560-0043, Japan}

\author{Naoki Koshimoto}
\affiliation{Department of Earth and Space Science, Graduate School of Science, Osaka University, Toyonaka, Osaka 560-0043, Japan}

\author{Nicholas J. Rattenbury}
\affiliation{Department of Physics, University of Auckland, Private Bag 92019, Auckland, New Zealand}

\author{Fumio Abe}
\affiliation{Institute for Space-Earth Environmental Research, Nagoya University, Nagoya 464-8601, Japan}
\author{Richard Barry}
\affiliation{Code 667, NASA Goddard Space Flight Center, Greenbelt, MD 20771, USA}
\author{David P.~Bennett}
\affiliation{Code 667, NASA Goddard Space Flight Center, Greenbelt, MD 20771, USA}
\affiliation{Department of Astronomy, University of Maryland, College Park, MD 20742, USA}
\author{Aparna Bhattacharya}
\affiliation{Code 667, NASA Goddard Space Flight Center, Greenbelt, MD 20771, USA}
\affiliation{Department of Astronomy, University of Maryland, College Park, MD 20742, USA}
\author{Akihiko Fukui}
\affiliation{Department of Earth and Planetary Science, Graduate School of Science, The University of Tokyo, 7-3-1 Hongo, Bunkyo-ku, Tokyo 113-0033, Japan}
\affiliation{Instituto de Astrofisica de Canarias, Via Lactea s/n, E-38205 La Laguna, Tenerife, Spain}
\author{Ryusei Hamada}
\affiliation{Department of Earth and Space Science, Graduate School of Science, Osaka University, Toyonaka, Osaka 560-0043, Japan}
\author{Shunya Hamada}
\affiliation{Department of Earth and Space Science, Graduate School of Science, Osaka University, Toyonaka, Osaka 560-0043, Japan}
\author{Naoto Hamasaki}
\affiliation{Department of Earth and Space Science, Graduate School of Science, Osaka University, Toyonaka, Osaka 560-0043, Japan}
\author{Yuki Hirao}
\affiliation{Institute of Astronomy, Graduate School of Science, The University of Tokyo, 2-21-1 Osawa, Mitaka, Tokyo 181-0015, Japan}
\author{Stela Ishitani Silva}
\affiliation{Department of Physics, The Catholic University of America, Washington, DC 20064, USA}
\affiliation{Code 667, NASA Goddard Space Flight Center, Greenbelt, MD 20771, USA}
\author{Yoshitaka Itow}
\affiliation{Institute for Space-Earth Environmental Research, Nagoya University, Nagoya 464-8601, Japan}
\author{Yutaka Matsubara}
\affiliation{Institute for Space-Earth Environmental Research, Nagoya University, Nagoya 464-8601, Japan}
\author{Shota Miyazaki}
\affiliation{Institute of Space and Astronautical Science, Japan Aerospace Exploration Agency, 3-1-1 Yoshinodai, Chuo, Sagamihara, Kanagawa 252-5210, Japan}
\author{Yasushi Muraki}
\affiliation{Institute for Space-Earth Environmental Research, Nagoya University, Nagoya 464-8601, Japan}
\author{Tsutsumi Nagai}
\affiliation{Department of Earth and Space Science, Graduate School of Science, Osaka University, Toyonaka, Osaka 560-0043, Japan}
\author{Greg Olmschenk}
\affiliation{Code 667, NASA Goddard Space Flight Center, Greenbelt, MD 20771, USA}
\author{Clement Ranc}
\affiliation{Sorbonne Universit\'e, CNRS, UMR 7095, Institut d'Astrophysique de Paris, 98 bis bd Arago, 75014 Paris, France}
\author{Yuki K. Satoh}
\affiliation{College of Science and Engineering, Kanto Gakuin University, 1-50-1 Mutsuurahigashi, Kanazawa-ku, Yokohama, Kanagawa 236-8501, Japan}
\author{Daisuke Suzuki}
\affiliation{Department of Earth and Space Science, Graduate School of Science, Osaka University, Toyonaka, Osaka 560-0043, Japan}
\author{Paul . J. Tristram}
\affiliation{University of Canterbury Mt.\,John Observatory, P.O. Box 56, Lake Tekapo 8770, New Zealand}
\author{Aikaterini Vandorou}
\affiliation{Code 667, NASA Goddard Space Flight Center, Greenbelt, MD 20771, USA}
\affiliation{Department of Astronomy, University of Maryland, College Park, MD 20742, USA}
\author{Hibiki Yama}
\affiliation{Department of Earth and Space Science, Graduate School of Science, Osaka University, Toyonaka, Osaka 560-0043, Japan}


\begin{abstract}
We present measurements of the microlensing optical depth and event rate toward the Galactic bulge using the dataset from the 2006--2014 MOA-II survey, which covers $22$ bulge fields spanning $\sim 42~\rm deg^2$ between $-5^\circ < l < 10^\circ$ and $-7^\circ < b < -1^\circ$.
In the central region with $|l|<5^\circ$, we estimate an optical depth of $\tau = [1.75\pm0.04]\times 10^{-6}\exp[{(0.34\pm0.02)(3^{\circ}-|b|)}]$ and an event rate of $\Gamma = [16.08\pm0.28]\times 10^{-6}\exp[{(0.44\pm0.02)(3^{\circ}-|b|)}] \rm star^{-1} year^{-1}$ using a sample consisting of $3525$ microlensing events, with Einstein radius crossing times of $t_{\rm E} < 760~\rm days$ and source star magnitude of $I_{\rm s}<21.4~\rm mag$.
We confirm our results are consistent with the latest measurements from OGLE-IV 8 year dataset \citep{mro19}.
We find our result is inconsistent with a prediction based on Galactic models, especially in the central region with $|b|<3^\circ$. 
These results can be used to improve the Galactic bulge model, and more central regions can be further elucidated by future microlensing experiments, such as The PRime-focus Infrared Microlensing Experiment (PRIME) and  Nancy Grace Roman Space Telescope.
\end{abstract}

\keywords{Milky Way Galaxy (1054), Gravitational microlensing (672)}

\section{Introduction} \label{sec-intro}
A gravitational microlensing enables us to test models of the structure, kinematics, and dynamics of our Galaxy through the measurement of event rates and optical depth as these are related to the mass, velocity, and density distributions in our Galaxy \citep{pac91,gri91}.
Currently, several groups: MOA-II\footnote{https://www.massey.ac.nz/~iabond/moa/alert2024/alert.php}, OGLE-IV\footnote{https://ogle.astrouw.edu.pl/ogle4/ews/ews.html}, and KMTNet\footnote{https://kmtnet.kasi.re.kr/ulens/}\citep{kim16} are conducting microlensing surveys toward the Galactic Bulge (GB) and detect a couple of thousand microlensing events every year.
In addition, the PRime-focus Infrared Microlensing Experiment (PRIME) began their survey toward the Galactic center in 2023 \citep{kon23,yam23}.

The magnification of a background source by microlensing is given by \citep{pac86}
\begin{equation}
A(u) = \frac{u^2 +2}{u\sqrt{u^2+4}},
\label{eq-amp} 
\end{equation}
where $u$ is the angular separation of the source and lens scaled by the angular Einstein radius,  $\theta_{\rm E}$, which is given by
\begin{equation}
\theta_{\rm E} = \sqrt{\kappa M_{\rm L} \left(\frac{\rm 1~au}{D_{\rm L}} - \frac{\rm 1~au}{D_{\rm S}}\right)},
\label{eq-thE} 
\end{equation}
where $\kappa = 8.144 ~{\rm mas} ~M_\odot ^{-1}$, $D_{\rm L}$ and $D_{\rm S}$ are distance to the lens and source, respectively, and $M_{\rm L}$ is the lens mass.
$u$ can be written as a function of time;
\begin{equation}
u(t) = \sqrt{u_{\rm 0}^2 + \left(\frac{t-t_{\rm 0}}{t_{\rm E}}\right)^2},
\label{eq-u} 
\end{equation}
where $u_{\rm 0}$, $t_{\rm 0}$, and $t_{\rm E}$ are the minimum impact parameter, the time of maximum magnification, and the Einstein radius crossing time, respectively.
The Einstein radius crossing time, $t_{\rm E}$, is given by $t_{\rm E} = \theta_{\rm E}/\mu_{\rm rel}$, where the lens-source relative proper motion $\mu_{\rm rel}$ is given by $\mu_{\rm rel} = |{\bm \mu_{\rm L}} - {\bm \mu_{\rm S}}|$ where ${\bm \mu_{\rm L}}$ is the lens proper motion vector, and ${\bm \mu_{\rm S}}$ is the source proper motion vector.

The microlensing optical depth, $\tau$, is defined as the probability that a given star is inside the Einstein ring of a lens at any given time. 
The probability that one source at distance, $D_{\rm S}$, is inside the Einstein ring, $\tau(D_{\rm S})$, is given by
\begin{equation}
\tau(D_{\rm S}) = \int^{D_{\rm S}}_{0}\Bigl[\int n(D_{\rm L},M_{\rm L}) \pi  R_{\rm E}^2(D_{\rm L},M_{\rm L}) dM_{\rm L}\Bigr] dD_{\rm L} = \pi \kappa \int^{D_{\rm S}}_{0} \rho(D_{\rm L}) D_{\rm L} \left(1 - \frac{D_{\rm L}}{D_{\rm S}}\right)  dD_{\rm L},
\label{eq-tau_ds} 
\end{equation}
where $n(D_{\rm L},M_{\rm L})$ is number density of lenses with the mass, $M_{\rm L}$, at the distance, $D_{\rm L}$, $R_{\rm E} = D_{\rm L} \theta_{\rm E}$ is the Einstein radius, and $\rho(D_{\rm L}) = \int M_{\rm L}  n(D_{\rm L},M_{\rm L}) dM_{\rm L}$ is the mass density of lenses. The observable quantity is just the integrated optical depth averaged over all detectable source stars toward the line of sight i.e.,
\begin{equation}
\tau = \frac{1}{N_{\rm s}} \int_{0}^{\infty} \tau(D_{\rm S}) N(D_{\rm S}) dD_{\rm S}
\label{eq-tau} 
\end{equation}
where $N(D_{\rm S})$ is the number density of source stars at distance, $D_{\rm S}$, and $N_{\rm s} = \int_{0}^{\infty} N(D_{\rm S}) dD_{\rm S}$ is number of all detectable source stars in the direction of sight.

The microlensing event rate is defined as the probability of an occurrence of a microlensing event per source star per unit time.
The area swept per unit time by the Einstein ring of a lens with mass, $M_{\rm L}$, and relative velocity $v_{\rm rel}$ at a distance $D_{\rm L}$ is given by $S(M_{\rm L}, D_{\rm L},v_{\rm rel}) = 2 v_{\rm rel} R_{\rm E}(M_{\rm L}, D_{\rm L}) ~(= 2 D_{\rm L}^2 \mu_{\rm rel} \theta_{\rm E}(M_{\rm L}, D_{\rm L}))$ \citep{bat11}.
The probability that one source at distance, $D_{\rm S}$, is microlensed in unit time, $\gamma$, can be calculated by integrating $S(M_{\rm L}, D_{\rm L},v_{\rm rel})$ i.e.,
\begin{equation}
\gamma(D_{\rm S}) = \int^{D_{\rm S}}_{0} dD_{\rm L}\int^{\infty}_{0} dM_{\rm L}\int^{\infty}_{0} d\mu_{\rm rel}~ S(M_{\rm L}, D_{\rm L},\mu_{\rm rel}) n(D_{\rm L},M_{\rm L},\mu_{\rm rel}),
\label{eq-gamma_ds} 
\end{equation}
where $n(D_{\rm L},M_{\rm L},\mu_{\rm rel})$ is the local number density of lenses with mass, $M_{\rm L}$, and relative proper motion, $\mu_{\rm rel}$, at distance, $D_{\rm L}$.
The microlensing event rate, $\Gamma$, is given by integration of $\gamma(D_{\rm S})$ with  number density of source stars, $N(D_{\rm S})$, i.e.,
\begin{equation}
\Gamma = \frac{1}{N_{\rm s}} \int_{0}^{\infty} \gamma(D_{\rm S}) N(D_{\rm S}) dD_{\rm S}.
\label{eq-gamma}
\end{equation}

The microlensing optical depth can also be expressed as the average ratio of the time when the source is within the Einstein radius to the duration of the survey.
For each microlensing event, the time that the source star is within the Einstein radius is given by $2t_{\rm E}\sqrt{1-u_{\rm 0}^2}$, however since the distribution in $u_{\rm 0}$ is uniform between $0 \leq u_{\rm 0} \leq 1$, we can use the average event duration which is given by $(\pi/2)t_{\rm E}$.
Thus, we can obtain the expression for $\tau$ as \citep{uda94}
\begin{equation}
\tau = \frac{\pi}{2 N_{\rm s} T_{\rm o}}\sum_{i=1}^{N_{\rm eve}}\frac{t_{\rm E,i}}{\epsilon(t_{\rm E,i})},
\label{eq-tau_obs} 
\end{equation}
where $N_{\rm eve}$ is number of observed events, $t_{\rm E,i}$ is Einstein radius crossing time of the $i$th event, $\epsilon(t_{\rm E,i})$  is the detection efficiency, $N_{\rm s}$ is the total number of monitored source stars, and $T_{\rm o}$ is the duration of the survey.

The microlensing event rate, $\Gamma$, can be more simply derived and expressed as
\begin{equation}
\Gamma = \frac{1}{N_{\rm s} T_{\rm o}}\sum_{i=1}^{N_{\rm eve}}\frac{1}{\epsilon(t_{\rm E,i})}.
\label{eq-gamma_obs} 
\end{equation}
This is the effective number of microlensing events, $N_{\rm eve, eff}\equiv \sum_{i=1}^{N_{\rm eve}}1/\epsilon(t_{\rm E, i})$, divided by the number of sources, $N_{\rm s}$ and the duration of the survey, $T_{\rm o}$.

The first measurement of the microlensing optical depth toward the GB was conducted by \citet{uda94} and they estimated $\tau = (3.3 \pm 1.2)\times 10^{-6}$ based on data from the first phase of OGLE survey during 1992--1993.
This is much more than the theoretical prediction of $\tau \sim 5 \times 10^{-7}$ \citep{pac91,gri91}, or $\tau \sim 8.5 \times 10^{-7}$ \citep{kir94}.
The result of $\tau \sim 3.9^{+1.8}_{-1.2}$ from MACHO \citep{alc97} also support such excesses.

To explain the excesses of the optical depth, the presence of a bar along the line of sight to the GB was suggested by \citet{pac94,zha95,zha96,pea98,gyu99}.
However, their predictions ranged over $\tau = 0.8-2.0 \times 10^{-6}$ and this is still below the observed value.

\citet{alc97} raised the possibility of a systematic bias in the optical depth measurement due to the degeneracy between $t_{\rm E}$ and $u_{\rm 0}$ in events with blended unresolved sources.
\citet{pop01} proposed that such a systematic bias can be eliminated using only events with bright source stars, such as red clump giants (RCGs).
Several measurements of the microlensing optical depth based on only events with bright source stars suggested lower optical depth: $\tau = 0.94 \pm 0.29 \times 10^{-6}$ \citep{afa03},  $\tau = 2.17^{+0.47}_{-0.38} \times 10^{-6}$ \citep{pop05}, and $\tau = (1.62 \pm0.23) \exp[(0.43\pm0.16)\rm deg^{-1}(|b|-3^\circ)]\times 10^{-6}$ \citep{ham06}.
However, their revised estimates were still larger than theoretical predictions.

Recent studies were conducted by MOA-II with 474 events \citep{sum13,sum16} and OGLE-IV with 8002 events \citep{mro19}.
\citet{sum13} estimated $\tau = (2.35 \pm0.18) \exp[(0.51\pm0.07)\rm deg^{-1}(|b|-3^\circ)]\times 10^{-6}$ based on 474 all-source events from two years of the MOA-II survey during 2006 -- 2007. 
In their analysis, the same luminosity function (LF) in all fields as the one in Baade’s window was used for estimating the number of sources.
\citet{sum16} pointed out the incompleteness of their number counts of red clump giants used to normalize that LF and they estimated $\tau = (1.84 \pm0.14) \exp[(0.44\pm0.07)\rm deg^{-1}(|b|-3^\circ)]\times 10^{-6}$ by correcting this incompleteness.
\citet{mro19} measured the optical depth using data from OGLE-IV sky survey from 2010 -- 2017 and their estimation, $\tau = (1.36 \pm0.04) \exp[(0.39\pm0.03)\rm deg^{-1}(|b|-3^\circ)]\times 10^{-6}$, is $\sim 30 \%$ smaller than the estimation form MOA-II \citep{sum16}.
They argue that the difference is probably caused by the wrong number of source stars used for calculations in \citet{sum16}.

Recently, \citet{kos23} systematically analyzed single-lens events from the MOA-II survey in 2006 -- 2014 for a measurement of the mass function of free-floating planets \citep{sum23}.
Their sample consists of $\sim$ 3500 events with $t_{\rm E} \leq 757 ~\rm days$. 
In this study, we present a measurement of the microlensing optical depth and event rate toward the GB, based on the microlensing sample from \citet{kos23}.

This paper is organized as follows.
Sec. \ref{sec-sample} presents the description of our data and the selection of microlensing events.
We present the microlensing optical depth and event rate results in Sec. \ref{sec-result}.
A comparison of our results with previous results and model predictions is described in Sec. \ref{sec-dis}.
A summary is in Sec. \ref{sec-conclu}.

\section{Data} \label{sec-sample}
The data set used in this analysis is the same one used in \citet{sum23}.
This was taken during the 2006--2014 seasons of the MOA-II high cadence photometric survey toward the Galactic bulge. 
MOA-II uses the $1.8\rm m$ MOA-II telescope located at the University of Canterbury’s Mount John Observatory in New Zealand.
The telescope is equipped with a wide field camera, MOA-cam3 \citep{sak08}, which consists of ten $2 \rm k \times 4 \rm k$ pixel CCDs with 15 $\mu$m pixels. With the pixel scale of $0.58 ~\rm {arcsec/pixel}$, this gives a 2.18 $\rm {deg}^2$ field of view (FOV).
The median seeing for this data set is $2.0\arcsec$.
The coordinates of the center of 20 MOA-II fields and the cadences are listed in Table \ref{table-field}.
Note that gb6 and gb22, and some subfields in other fields, in all about $12\%$ of the total survey area, are not used in our analysis because an RCG population could not be identified clearly in the CMD.
Each field is divided into 80 subfields in which the photometric scale is individually calibrated in each subfield.
The numbers of subfields for each field are also given in Table \ref{table-field}.
The total duration of the dataset is 3146 days over the period $\rm HJD = 2453824-2456970$.

Data analysis and microlensing event selection for this data set were conducted by \citet{kos23}, which is a companion paper of \citet{sum23}.
Here, we briefly introduce their process, but for more details, see \citet{kos23}.

\begin{deluxetable*}{cccccccccc}
    \tablewidth{10pt} 
    \tablenum{1}
    \tablecaption{MOA-II Galactic Bulge Fields with Galactic Coordinates of the Mean Field Center ($\langle l \rangle$,$\langle b \rangle$), the Number of Subfields Used ($N_{\rm sub}$), the Number of Source Stars ($N_{\rm s}$), the Number of Microlensing Events ($N_{\rm ev}$), the Microlensing Event Rate per Star per Year ($\Gamma$), the Microlensing Event Rate per Square Degree per Year ($\Gamma_{\rm deg^2}$), the Optical Depth ($\tau$), and the average Einstein radius crossing time $\langle t_{\rm E} \rangle$. \label{table-field}}
    \tablehead{
    \colhead{Field} & \colhead{$\langle l \rangle$} & \colhead{$\langle b \rangle$} & \colhead{$N_{\rm sub}$} & \colhead{$N_{\rm s}$} & \colhead{{$N_{\rm eve}$}} & \colhead{$\tau$} & \colhead{$\Gamma$} & \colhead{$\Gamma_{\rm deg^2}$} & \colhead{$\langle t_{\rm E}\rangle$}\\
    \colhead{} & \colhead{($^\circ$)} & \colhead{($^\circ$)} & \colhead{} & \colhead{} & \colhead{} & \colhead{$(10^{-6}$)} & \colhead{$(10^{-6}\rm star ^{-1} year ^{-1})$} & \colhead{$(\rm deg ^{-2} year ^{-1})$} & \colhead{$(\rm days)$}
    }
    \startdata 
    $\rm {gb}1$ & $-4.33$ & $-3.11$ & $79$ & $21047010$ & $193$ &  $1.68^{+0.15}_{-0.15}$ & $12.86^{+0.27}_{-0.27}$ & $125.73^{+2.69}_{-2.65}$ & $29.47{\pm1.12}$ \\
    $\rm {gb}2$ & $-3.86$ & $-4.39$ & $79$ & $17647488$ & $133$ &  $1.23^{+0.14}_{-0.13}$ & $8.94^{+0.25}_{-0.25}$ & $73.33^{+2.06}_{-2.02}$ & $31.22{\pm1.30}$ \\
    $\rm {gb}3$ & $-2.35$ & $-3.51$ & $79$ & $22711037$ & $193$ &  $1.38^{+0.13}_{-0.11}$ & $11.30^{+0.25}_{-0.24}$ & $119.20^{+2.62}_{-2.58}$ & $27.53{\pm1.03}$ \\
    $\rm {gb}4$ & $-0.83$ & $-2.63$ & $77$ & $25985143$ & $307$ &  $1.98^{+0.15}_{-0.14}$ & $20.46^{+0.31}_{-0.31}$ & $253.43^{+3.85}_{-3.81}$ & $21.91{\pm0.70}$ \\
    $\rm {gb}5$ & $0.65$ & $-1.86$ & $65$ & $29137851$ & $492$ &  $3.44^{+0.22}_{-0.21}$ & $31.43^{+0.36}_{-0.36}$ & $516.98^{+5.98}_{-5.94}$ & $24.77{\pm0.77}$ \\
    $\rm {gb}7$ & $-1.72$ & $-4.60$ & $78$ & $16344845$ & $99$ &  $0.81^{+0.09}_{-0.08}$ & $7.91^{+0.25}_{-0.24}$ & $60.86^{+1.89}_{-1.85}$ & $23.18{\pm0.83}$ \\
    $\rm {gb}8$ & $-0.19$ & $-3.75$ & $78$ & $22263658$ & $186$ &  $1.37^{+0.14}_{-0.12}$ & $12.73^{+0.27}_{-0.26}$ & $133.39^{+2.78}_{-2.74}$ & $24.39{\pm1.05}$ \\
    $\rm {gb}9$ & $1.33$ & $-2.88$ & $79$ & $33308780$ & $466$ &  $2.19^{+0.14}_{-0.13}$ & $21.63^{+0.28}_{-0.28}$ & $334.69^{+4.37}_{-4.33}$ & $22.86{\pm0.67}$ \\
    $\rm {gb}10$ & $2.84$ & $-2.09$ & $70$ & $21465124$ & $282$ &  $2.56^{+0.28}_{-0.26}$ & $22.81^{+0.36}_{-0.36}$ & $256.71^{+4.07}_{-4.03}$ & $25.35{\pm1.39}$ \\
    $\rm {gb}11$ & $-1.11$ & $-5.73$ & $76$ & $10931979$ & $46$ &  $0.59^{+0.11}_{-0.10}$ & $4.80^{+0.24}_{-0.23}$ & $25.33^{+1.24}_{-1.20}$ & $28.02{\pm2.04}$ \\
    $\rm {gb}12$ & $0.44$ & $-4.87$ & $79$ & $16090446$ & $86$ &  $0.79^{+0.10}_{-0.10}$ & $7.06^{+0.23}_{-0.23}$ & $52.76^{+1.75}_{-1.71}$ & $25.46{\pm1.20}$ \\
    $\rm {gb}13$ & $1.97$ & $-4.02$ & $79$ & $23728248$ & $188$ &  $1.32^{+0.13}_{-0.11}$ & $11.03^{+0.24}_{-0.24}$ & $121.53^{+2.64}_{-2.60}$ & $27.08{\pm1.08}$ \\
    $\rm {gb}14$ & $3.51$ & $-3.17$ & $79$ & $25094851$ & $254$ &  $1.58^{+0.14}_{-0.13}$ & $14.33^{+0.27}_{-0.26}$ & $166.99^{+3.09}_{-3.05}$ & $24.95{\pm0.90}$ \\
    $\rm {gb}15$ & $4.99$ & $-2.45$ & $62$ & $10404096$ & $82$ &  $1.86^{+0.29}_{-0.25}$ & $13.02^{+0.39}_{-0.39}$ & $80.18^{+2.43}_{-2.38}$ & $32.34{\pm2.04}$ \\
    $\rm {gb}16$ & $2.60$ & $-5.17$ & $79$ & $15028199$ & $99$ &  $1.17^{+0.17}_{-0.16}$ & $7.38^{+0.25}_{-0.24}$ & $51.49^{+1.73}_{-1.69}$ & $35.81{\pm2.28}$ \\
    $\rm {gb}17$ & $4.15$ & $-4.34$ & $79$ & $17979175$ & $138$ &  $1.09^{+0.11}_{-0.11}$ & $9.37^{+0.25}_{-0.25}$ & $78.28^{+2.12}_{-2.09}$ & $26.20{\pm0.97}$ \\
    $\rm {gb}18$ & $5.69$ & $-3.51$ & $78$ & $15478756$ & $104$ &  $1.27^{+0.17}_{-0.15}$ & $8.56^{+0.26}_{-0.26}$ & $62.37^{+1.91}_{-1.87}$ & $33.50{\pm1.81}$ \\
    $\rm {gb}19$ & $6.54$ & $-4.57$ & $78$ & $12087586$ & $81$ &  $1.15^{+0.16}_{-0.14}$ & $7.48^{+0.28}_{-0.27}$ & $42.56^{+1.58}_{-1.54}$ & $34.81{\pm1.95}$ \\
    $\rm {gb}20$ & $8.10$ & $-3.75$ & $79$ & $10730219$ & $65$ &  $1.19^{+0.21}_{-0.17}$ & $7.25^{+0.29}_{-0.28}$ & $36.14^{+1.45}_{-1.41}$ & $37.27{\pm2.56}$ \\
    $\rm {gb}21$ & $9.60$ & $-2.94$ & $73$ & $8321495$ & $31$ &  $0.70^{+0.16}_{-0.14}$ & $4.87^{+0.27}_{-0.26}$ & $20.36^{+1.14}_{-1.10}$ & $32.68{\pm2.79}$ \\
    \hline
    $\rm all$ & $1.85$ & $-3.69$ & $1536$ & $378451020$ & $3525$ &  $1.61^{+0.04}_{-0.04}$ & $14.00^{+0.07}_{-0.07}$ & $126.60^{+0.61}_{-0.61}$ & $26.06{\pm0.27}$ \\
    \enddata
    \end{deluxetable*}

\subsection{Data Analysis}\label{sec-data}
The observed images were reduced with the MOA implementation \citep{bon01} of the difference image analysis (DIA) method \citep{tom96,ala96,ala00}.
This makes it possible to perform precise photometry even in very crowded fields such as the GB.
Each field consists of 10 chips and each chip is divided into eight $1024\times1024$ pixel subfields during the DIA process.
All photometric light curves were de-trended by fitting a polynomial model given by Eq. 4 of \citet{kos23} to correct systematic trends correlated with seeing and airmass.
The DIA light curve photometry values are given as flux values which are scaled to the MOA reference images.
Calibration for MOA reference images was performed by cross-referencing the MOA-II DOPHOT catalog to the OGLE-III photometry map of the Galactic bulge \citep{szy11}.

\subsection{Microlensing Event Selection}\label{sec-selec}
First, \citet{kos23} detected variable objects on the subtracted images by using a custom implementation of the IRAF task DAOFIND \citep{ste87} with the modification that both positive and negative PSF profiles are searched for simultaneously.
They eliminated the detection of spurious variations among all detected variable objects using the criteria listed in Table 2 of \citet{kos23}.
As a result, $2,409,061$ variable objects were detected at this stage of the analysis.

Second, they created the light curves of variable objects by using PSF-fitting photometry on the difference images.
By imposing criteria on the number of data points and the signal-to-noise ratio of the light curve, $67,242$ light curves were selected and $6,111$ microlensing candidates were found during this process.

Finally, they selected high-quality single-lens events based on the uncertainty of Einstein radius crossing times, the number of data points in the peak, and the residual from the best-fit model, etc.
As a result, $3,554$ and $3,535$ objects remained as microlensing candidates after applying nominal criteria (CR1) and stricter criteria (CR2), respectively, among all visually identified $6,111$ candidates. 
All criteria are summarized in Table 2 of \citet{kos23}.
In our analysis, we use only a sample from CR2, but in addition to the CR2 cut, we remove events with $t_{\rm E} \leq 1 ~\rm days$ because such short-timescale events are regarded as free-floating planet candidates which are not considered in predictions of microlensing optical depth and event rate from Galactic models.

Due to these selection criteria, all events with significant binary lens features were removed from the sample.
To correct our optical depth measurement for binary lens events excluded from the sample, we assume that the fraction of binary lens events among all microlensing events is $6\%$.
As described in \citet{sum13}, this correction can be achieved by applying a factor of 1.09 to the optical depth and 1.06 to the event rate.
The number of events in each field is listed in Table \ref{table-field}.

\subsection{Detection Efficiency}\label{sec-eff}
For the calculation of detection efficiency, \citet{kos23} performed image-level simulations of $6.4 \times 10^7$ artificial events.
They generated $40,000$ artificial events in each subfield and embedded them at random positions between $0 \leq x \rm{/pix} \leq 2048$ and $0 \leq y \rm{/pix} \leq 4096$ in each CCD. 
The microlensing parameters are randomly assigned between $3824 \leq t_{0}\rm{/JD^\prime} \leq 6970$, $0 \leq u_{\rm 0} \leq 1.5$, and a source magnitude of $14.2 \leq I_{\rm s}\rm/mag \leq 22$, uniformly.
The detection efficiency was calculated as a function of Einstein radius crossing time, $t_{\rm E}$, and Einstein radius, $\theta_{\rm E}$, in each subfield.

We use the average detection efficiency of each field which is derived by integrating the detection efficiency of all subfields within that field weighted by $n_{\mathrm{RC},k}^2 f_{\mathrm{LF},k}(I_{\mathrm{s}, i})$, where $n_{\mathrm{RC},k}$ is the number density of RCGs in the $k$th subfield, $f_{\mathrm{LF},k}$ is the fraction of stars that have a source magnitude of $i$th artificial event, $I_{\mathrm{s}, i}$, given by the luminosity function (LF) in $k$th subfield (Eq. 12, 13 of \citet{kos23}).

In our analysis, we use the one-dimensional detection efficiency as a function of $t_{\rm E}$ (see Eq. \ref{eq-tau_obs}, \ref{eq-gamma_obs}).
This is calculated by integrating the 2D detection efficiency, $\epsilon(t_{\rm E}, \theta_{\rm E})$, over $\theta_{\rm E}$ weighted by the fraction of events with $\theta_{\rm E}$ among events with $t_{\rm E}$ in the model, $\Gamma(\theta_{\rm E} | t_{\rm E})$, (Eq.11 of \citet{kos23}).
The integrated detection efficiency for each field is illustrated in Fig 4 of \citet{sum23}.

\subsection{Star Counts}\label{sec-ns}
For source star counts, we use a method similar to that of \citet{sum16}, but with some updates.

\citet{sum16} constructed a combined luminosity function (LF) in Baade's window by using the MOA-II Dophot star catalog for bright stars, and Hubble Space Telescope (HST) deep imaging \citep{hol98} for faint stars down to $I = 24~\rm mag$.
On the other hand, in this work, we used the LF based on the OGLE-III photometry map \citep{szy11} for bright stars.
Because the OGLE LF is more accurate and deeper than the MOA-II LF due to superior seeing, 
it can be accurately normalized and aligned by using magnitude ranges overlapping the brighter end of the HST LF, 
This improved the accuracy of the number count of the source stars.
We used RCG to calibrate and normalize this combined LF to the extinction and GB distance for each subfield because RCG stars serve as reliable standard candles \citep{kir97,sta00} and a number of which can be considered to be promotional to the number of source stars in each subfield.
We estimated the $I-\rm band$ magnitude of the center of RC, $I_{\rm RC}$, and the number of RCG stars, $N_{\rm RC}$ by fitting the LF of the MOA's reference images in each subfield with Eq. (4) of \citet{nat13}.
\citet{sum16} investigated the completeness of $N_{\rm RC}$ in the MOA-II GB fields by comparing it with that of OGLE-III \citep{nat13}, and found that the ratio of $N_{\rm RC}$ between MOA-II and OGLE-III, $f_{\rm RC}$, is well expressed by $f_{\rm RC} = (0.62\pm0.01)-(0.052\pm0.003)\times b$ (Eq. (2) of \citet{sum16}).
We assume that $N_{\rm RC}$ of OGLE-III is complete and correct the incompleteness of our $N_{\rm RC}$ using this linear relationship.
We calculate the number of source stars, $N_{\rm s}$, by integrating this scaled–combined LF over the specified magnitude range (i.e., $10\leq I_{\rm s} \leq 21.4$).

We also made two other updates in the calibration of the LF.
As we mentioned above, we used $I_{\rm RC}$ and $N_{\rm RC}$ for calibrating the LF. 
We derived $I_{\rm RC}$ and $N_{\rm RC}$ by fitting the magnitude distribution of the reference images in each subfield with Eq. (4) of \citet{nat13}. 
\citet{sum16} ignored the term related to the asymptotic giant branch bump (AGBB) in Eq. (4) of \citet{nat13}. 
Since this omission may have introduced systematic errors in the calculation of $I_{\rm RC}$ and $N_{\rm RC}$, in this work, we incorporated this term into the calculation.
We also updated the calibration of the photometric zero-point from instrumental magnitudes of the MOA reference images to the Kron/Cousins $I$ band, which is also important for calibrating the LF.
\citet{sum16} estimated the mean magnitude zero-point from the $30\%$ of MOA-II fields that overlap with the OGLE-II map and applied this mean zero-point to all of the fields.
This approach, however, introduced an uncertainty of approximately $0.25~\rm mag$ in the calibration of each subfield.
Thus, we instead use the OGLE-III photometry map to calibrate for each
 subfield individually, which provides more precise calibration. 
These updates improved the accuracy of the source star counts in this study.

As a result of these updates, our source star counts are a factor of $1.5$ larger than the previous source star counts by \citet{sum16} as described in Sec. \ref{sec-comp_moa}.
In addition, we compared the number densities of stars brighter than $I=21~\rm mag$ calculated by this updated method with those from \citet{mro19} in each subfield and found that they are consistent within a factor of $1.06^{+0.39}_{-0.25} (= \rm MOA/OGLE)$.

\begin{figure*}
\centering
\includegraphics[width=19cm]{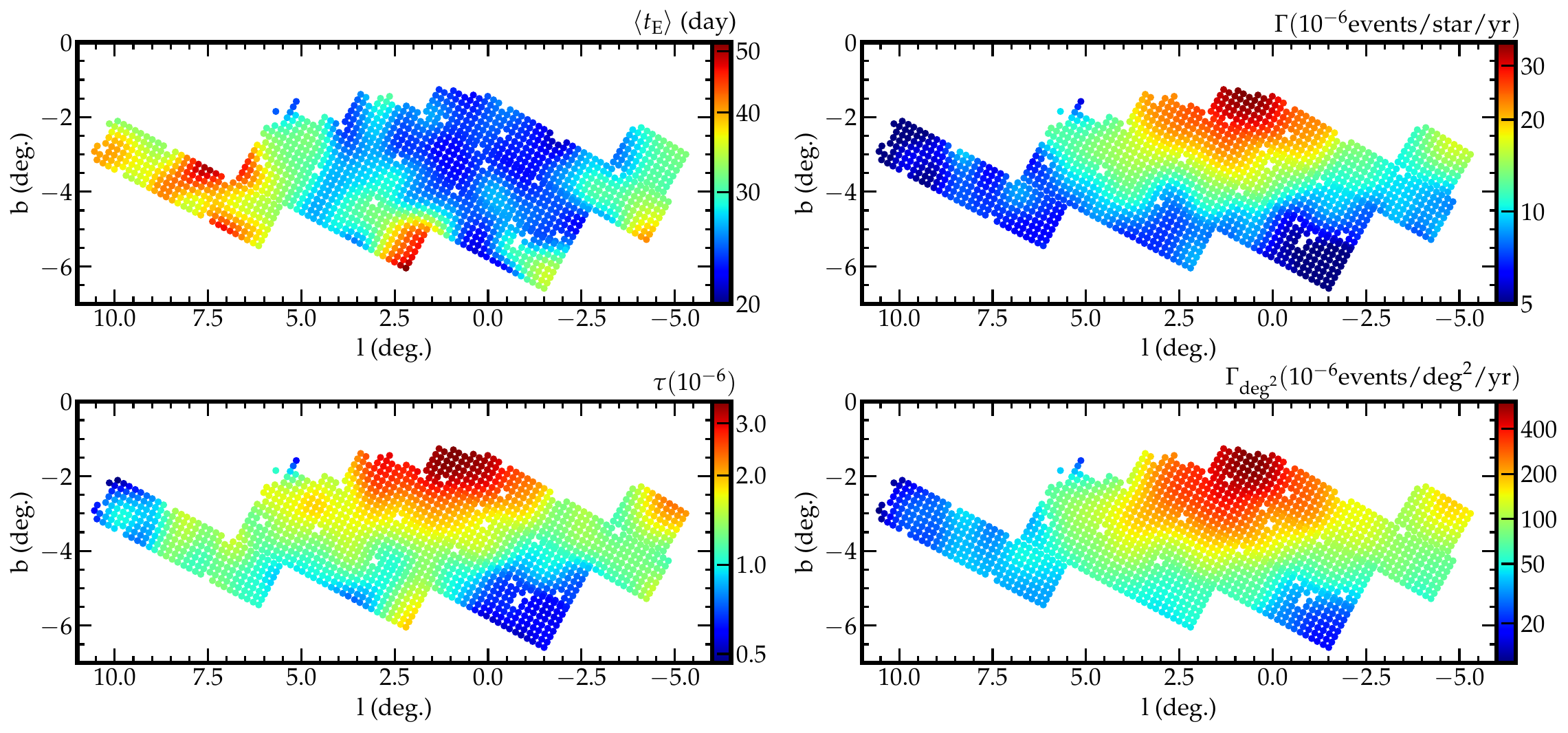}
\caption{Color map of the mean Einstein radius crossing time, $\langle t_{\rm E} \rangle$ (top left), the microlensing optical depth, $\tau$, (bottom left), and the event rate per tar per year, $\Gamma$, (top right), the event rate per square degree per year, $\Gamma_{\deg^{2}}$, (bottom right) based on the MOA-II microlensing survey in 2006 -- 2014.
All maps are smoothed with a Gaussian with $\sigma = 24\arcmin$ within $1^\circ$ around the subfield.
\label{fig-map1}}
\end{figure*}

\begin{deluxetable*}{ccccccc}
    \tablewidth{20pt} 
    \tablenum{2}
    \tablecaption{The microlensing opetical depth and event rate for source with $I < 21.4$ in $b$ within $|l| < 5^\circ$.\label{table-result01}}
    \tablehead{
    \colhead{$\langle b \rangle$} & \colhead{$N_{\rm sub}$} & \colhead{$N_{\rm s}$} & \colhead{{$N_{\rm eve}$}} & \colhead{$\tau$} & \colhead{$\Gamma$} & \colhead{$\Gamma_{\rm deg^2}$}\\
    \colhead{($^\circ$)} & \colhead{} & \colhead{} & \colhead{} & \colhead{$(10^{-6}$)} & \colhead{$10^{-6}\rm star ^{-1} year ^{-1}$} & \colhead{$\rm deg ^{-2} year ^{-1}$}
    }
    \startdata 
    $-1.4012$ & $20$ & $4595283$ & $57$ &  $2.19^{+0.43}_{-0.34}$ & $31.70^{+4.38}_{-4.02}$ & $267.25^{+36.95}_{-33.86}$\\
    $-1.7690$ & $70$ & $25837259$ & $395$ &  $3.07^{+0.23}_{-0.21}$ & $30.84^{+1.57}_{-1.52}$ & $417.74^{+21.27}_{-20.57}$\\
    $-2.2645$ & $114$ & $38087885$ & $540$ &  $2.68^{+0.19}_{-0.18}$ & $24.61^{+1.07}_{-1.04}$ & $301.70^{+13.10}_{-12.73}$\\
    $-2.7576$ & $146$ & $50452070$ & $585$ &  $1.91^{+0.11}_{-0.10}$ & $18.31^{+0.76}_{-0.74}$ & $232.21^{+9.68}_{-9.42}$\\
    $-3.2486$ & $168$ & $54692934$ & $567$ &  $1.70^{+0.10}_{-0.09}$ & $15.43^{+0.65}_{-0.64}$ & $184.39^{+7.81}_{-7.60}$\\
    $-3.7490$ & $172$ & $49078606$ & $400$ &  $1.31^{+0.08}_{-0.08}$ & $11.51^{+0.58}_{-0.56}$ & $120.50^{+6.10}_{-5.90}$\\
    $-4.2512$ & $172$ & $42129953$ & $295$ &  $1.12^{+0.09}_{-0.08}$ & $9.38^{+0.55}_{-0.53}$ & $84.29^{+4.98}_{-4.79}$\\
    $-4.7410$ & $154$ & $32367770$ & $206$ &  $1.00^{+0.09}_{-0.08}$ & $8.18^{+0.58}_{-0.55}$ & $63.09^{+4.48}_{-4.28}$\\
    $-5.2270$ & $101$ & $18008997$ & $111$ &  $0.92^{+0.12}_{-0.10}$ & $7.59^{+0.74}_{-0.70}$ & $49.65^{+4.85}_{-4.55}$\\
    $-5.7197$ & $56$ & $8310438$ & $38$ &  $0.84^{+0.20}_{-0.16}$ & $5.24^{+0.90}_{-0.81}$ & $28.54^{+4.89}_{-4.40}$\\
    $-6.1945$ & $19$ & $2428901$ & $9$ &  $0.43^{+0.19}_{-0.14}$ & $4.00^{+1.52}_{-1.23}$ & $18.76^{+7.13}_{-5.76}$\\
    \enddata
    \tablecomments{ $\langle b \rangle$, $N_{\rm sub}$, $N_{\rm s}$, and $N_{\rm eve}$ indicate the average Galactic latitude, number of subfields, number of source stars, and number of microlensing events, respectively. }
    \end{deluxetable*}

\section{Result}\label{sec-result}
In this section, we estimated the optical depth and the event rate by following \citet{sum13} and \citet{sum16}.
We calculated the optical depth, $\tau$, the event rate per star per year, $\Gamma$, and the event rate per square degree per year, $\Gamma_{\deg^{2}}$, in each subfield by using Eq. \ref{eq-tau_obs} and \ref{eq-gamma_obs} and show the results in Fig. \ref{fig-map1}.
In addition, the mean Einstein radius crossing time weighted by detection efficiency $(= N_{\rm eve, eff}^{-1}\sum_{i=1}^{N_{\rm eve}}t_{\rm E, i}/\epsilon(t_{\rm E, i}))$, is also presented.
Note that all maps are smoothed with a Gaussian with $\sigma = 24\arcmin$ within $1^\circ$ around the subfield.
The combined average optical depth and event rate in each field are listed in Table \ref{table-field}.
Uncertainties in the event rates are calculated by Poisson statistics, but a calculation of uncertainty in the optical depth is more complicated because it does not follow Poisson statistics.
To estimate the uncertainty in the optical depth, we follow the bootstrap Monte Carlo method described in Section 6.1 of \citet{alc97}.
Regarding the mean Einstein radius crossing time, $\langle t_{\rm E} \rangle$, we present the standard errors of the mean. 

We estimated the average optical depth in all fields combined as $\tau = (1.61{\pm0.04})\times 10^{-6}$.
This is $\sim 5 \%$ higher than the result of $\tau = 1.53^{+0.12}_{-0.11}\times 10^{-6}$ based on the first two years of MOA-II survey \citep{sum16} but note that their result is based on source stars with $I<20$ whereas ours is based on source stars with $I<21.4$.
A comparison in the same magnitude range is made in Sec. \ref{sec-comp_moa}.
The measured optical depth and event rate of the central region with $l < 5^\circ$ is shown in Table \ref{table-result01}.
This is binned with a bin width of $\Delta b = 30\arcmin$ between $-6.5^\circ < b < -1^\circ$. 

\begin{figure}
\centering
\includegraphics[width=10cm]{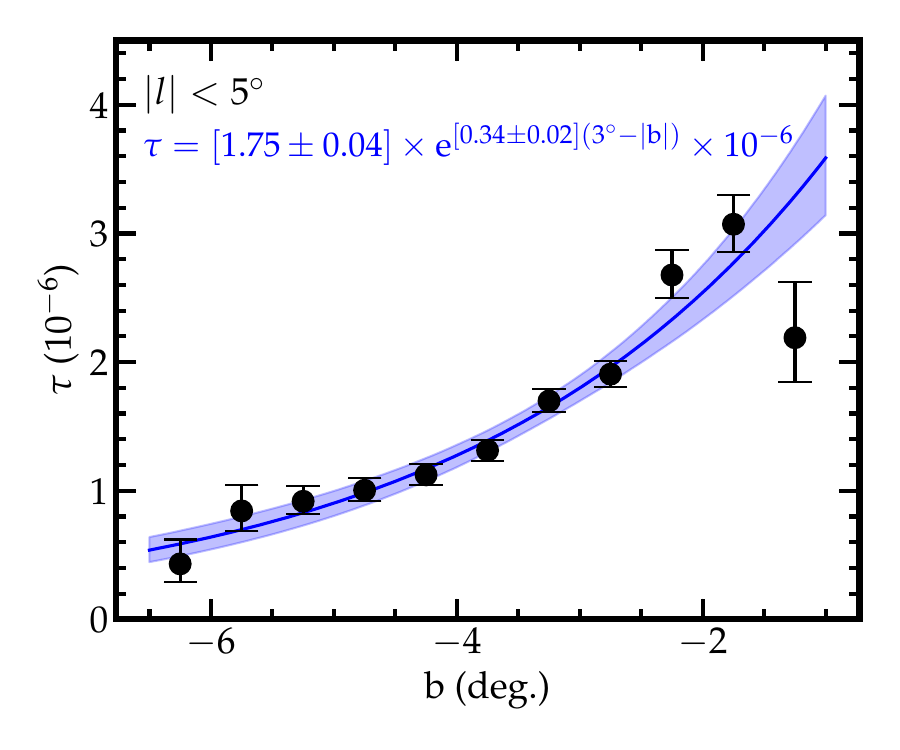}
\caption{Measured optical depths for sources with $I<21.4$ as a function of Galactic latitude $b$ within $|l| < 5^\circ$ (filled circles).
The values of subfields are binned with a width of $30\arcmin$.
Each value of the data is listed in Table \ref{table-result01}.
The solid lines and shaded area indicate the median and $99\%$ confidence level of the MCMC results, respectively.
\label{fig-tau_l<5}}
\end{figure}

\subsection{Fitting the Optical Depth with Parametric Model}\label{sec-fit_tau}
In this section, we present a fitting with a parametric model to the result of the optical depth in Table \ref{table-result01}.
To estimate the Probability Density Function (PDF) of the measured optical depth in each bin, we resample $N$ events from the sample within each bin, where $N$ is drawn from a Poisson distribution with a parameter equal to the number of events in each bin.
Assuming that the optical depth values computed in each subsample follow the PDF of measured optical depth, we estimate it by performing Kernel Density Estimation using that set of optical depth.

We adopted the simple exponential model given by
$\tau = \tau_{0} \exp[{c_{\tau}(3^{\circ}-|b|)}]$ and found $\tau_{0} = (1.75\pm0.04)\times 10^{-6}$ and $c_{\tau} = 0.34\pm0.02$ at $|l| < 5^\circ$ by using the Markov Chain Monte Carlo (MCMC) method.
The median and $99 \%$ confidence level of the MCMC results and data are shown in Fig. \ref{fig-tau_l<5}.

\begin{figure}
\centering
\includegraphics[width=10cm]{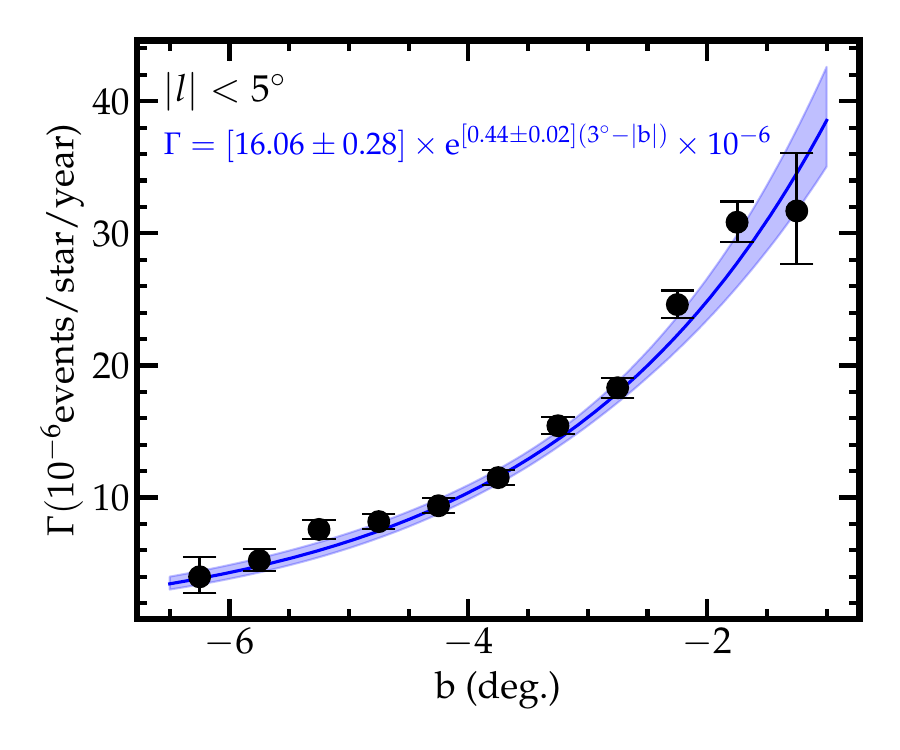}
\caption{Measured event rate for sources with $I<21.4$ as a function of Galactic latitude $b$ within $|l| < 3^\circ$ (filled circles).
The values of subfields are binned with a width of $30\arcmin$.
Each value of the data is listed in Table \ref{table-result01}.
The solid lines and shaded area indicate the median and $99\%$ confidence level of the MCMC results, respectively.
\label{fig-gamma_l<5}}
\end{figure}

\subsection{Fitting the Event Rate with Parametric Model}\label{sec-fit_gam}
Modeling the event rate is much simpler than that in the optical depth because the event rate follows Poisson statistics.
From the definition of the event rate, The expected number of events, $N_{\rm exp}$, under an assumed model at coordinates $(l,b)$, $\Gamma_{\rm model}(l,b)$, is given by
\begin{equation}
N_{\rm exp}(l,b;\Gamma_{\rm model}) = \Gamma_{\rm model}(l,b)N_{\rm s}(l,b) T_{\rm o} \langle \epsilon(l,b) \rangle,
\label{eq-nexp} 
\end{equation}
where $N_{\rm s}(l,b)$ and $\langle \epsilon(l,b) \rangle$ are number of source stars and detection efficiency averaged over $t_{\rm E}$ weighted by $t_{\rm E}$ distribution at coordinates $(l,b)$, respectively.
For calculating the $t_{\rm E}$ distribution at coordinates $(l,b)$, we use the Galactic model developed by \citet{kos21} and their microlensing event simulation tool, \texttt{genulens}\footnote{\url{https://github.com/nkoshimoto/genulens}} \citep{kos22}.
$N_{\rm s}^{\rm eff}(l,b)\equiv N_{\rm s}(l,b)\langle \epsilon(l,b) \rangle$ can be regarded as the effective number of source stars, taking into account the detection efficiency.
We denote $f$ as $\Gamma_{\rm model}T_{\rm o}$, which can be interpreted as the probability of a single star being microlensed during duration, $T_{\rm o}$.
Under the effective number of stars, $N_{\rm s}^{\rm eff}$, and microlensed probability, $f$, the probability of the number of events, $N_{\rm eve}$, is given by
\begin{equation}
P[N_{\rm eve}] = \binom{N_{\rm s}^{\rm eff}}{N_{\rm eve}}f^{N_{\rm eve}}(1-f)^{N_{\rm s}^{\rm eff}-N_{\rm eve}} \xrightarrow{N_{\rm s}^{\rm eff} \to \infty} \frac{(f N_{\rm s}^{\rm eff})^{N_{\rm eve}}}{N_{\rm eve}!}e^{-f N_{\rm s}^{\rm eff}}.
\label{eq-p_neve} 
\end{equation}
Note that the notations for $(l,b)$ are omitted in this equation.\\
Because $f(l,b) N^{\rm eff}_{\rm s}(l,b)= N_{\rm exp}(l,b)$ (Eq. \ref{eq-nexp}), we can finally derive the expression for the likelihood as
\begin{equation}
{\cal L}\left[N_{\rm exp}(l,b;\Gamma_{\rm model})|N_{\rm eve}(l,b)\right] = \frac{N_{\rm exp}(l,b)^{N_{\rm eve}(l,b)}}{N_{\rm eve}(l,b)!}e^{-N_{\rm exp}(l,b)}.
\label{eq-gam_like} 
\end{equation}
Under a uniform prior, a model can be evaluated by calculating $\prod_{(l,b)}{\cal L}\left[N_{\rm exp}(l,b;\Gamma_{\rm model})|N_{\rm eve}(l,b)\right]$.

As well as the case of optical depth, we adopt the simple exponential model given by $\Gamma = \Gamma_{0} \exp[{c_{\Gamma}(3^{\circ}-|b|)}]$.
Using the MCMC method, we estimated that $\Gamma_{0} = (16.08\pm 0.28)\times 10^{-6}$ and $c_{\Gamma} = 0.44\pm 0.02$.
The median and $99 \%$ confidence level of the MCMC result and the data are shown in Fig. \ref{fig-gamma_l<5}.

\begin{figure*}
\centering
\includegraphics[width=10cm]{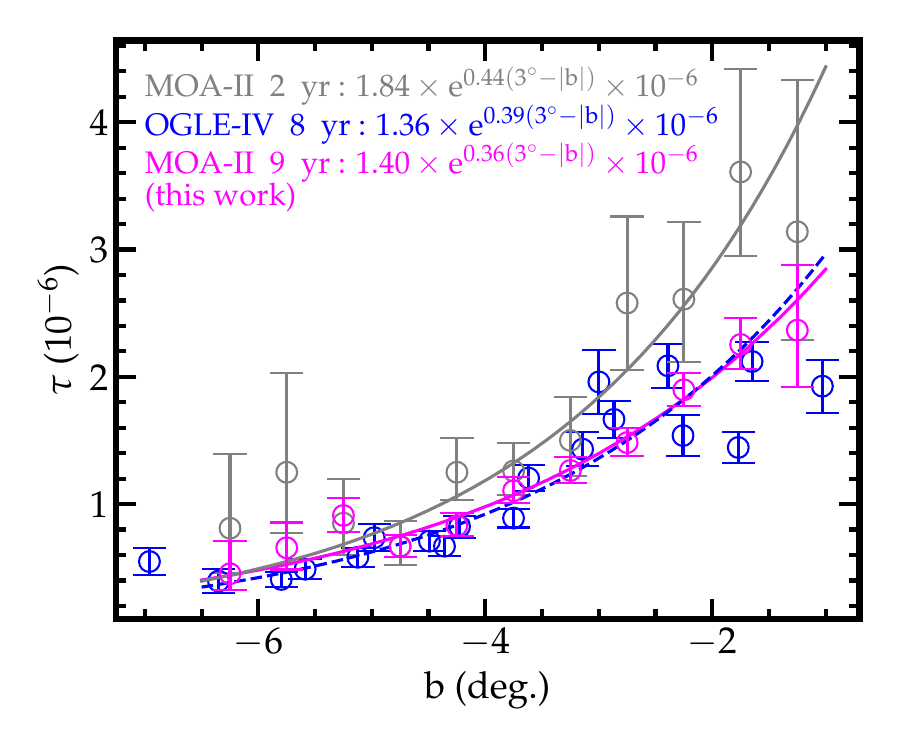}
\caption{
Comparison of microlensing optical depth of the central region $(|l|<3^\circ)$ between MOA-II $9 \rm yr$ (this work), MOA-II $2 \rm yr$ \citep{sum16} and OGLE-IV $8 \rm yr$ \citep{mro19}.
These measurements are based on the events with $I_{\rm s} < 20 ~\rm mag$ and $t_{\rm E} < 200~\rm days$.
The gray, blue, and magenta circles indicate the measurements by MOA-II $2 \rm yr$, OGLE-IV $8 \rm yr$, and MOA-II $9 \rm yr$, respectively.
The corresponding solid lines in each color represent the best-fit models for these datasets.
\label{fig-tau_comp}}
\end{figure*}

\begin{figure*}
\centering
\includegraphics[width=10cm]{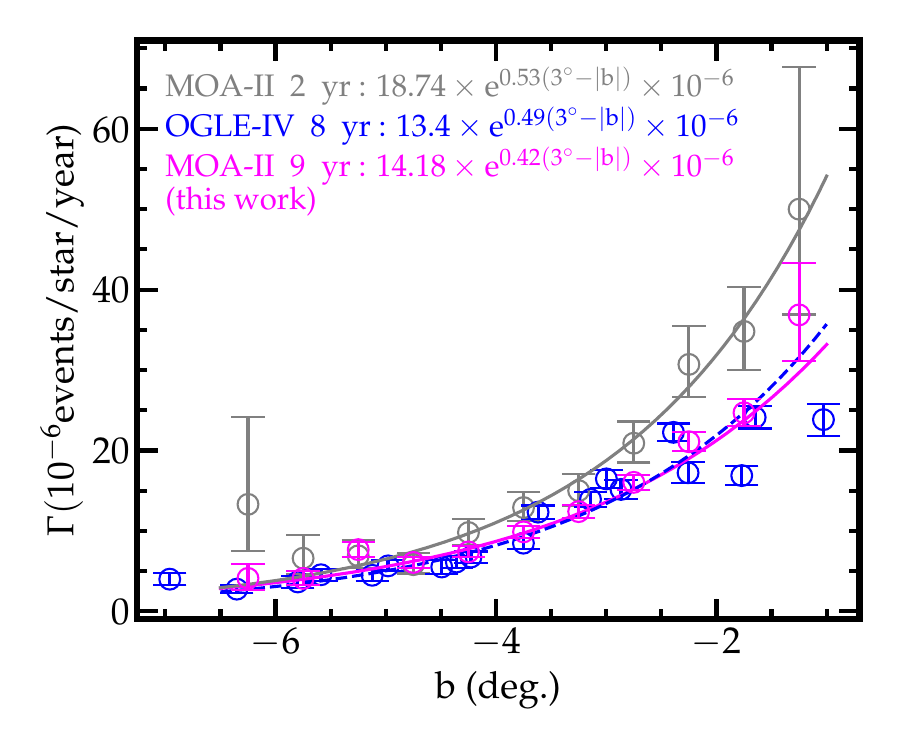}
\caption{
Comparison of microlensing event rate of the central region $(|l|<3^\circ)$ between MOA-II $9 \rm yr$ (this work), MOA-II $2 \rm yr$ \citep{sum16} and OGLE-IV $8 \rm yr$ \citep{mro19}.
These measurements are based on the events with $I_{\rm s} < 20 ~\rm mag$ and $t_{\rm E} < 200~\rm days$.
The gray, blue, and magenta circles indicate the measurements by MOA-II $2 \rm yr$, OGLE-IV $8 \rm yr$, and MOA-II $9 \rm yr$, respectively.
The corresponding solid lines in each color represent the best-fit models for these datasets.
\label{fig-gamma_comp}}
\end{figure*}

\begin{figure*}
\centering
\includegraphics[width=10cm]{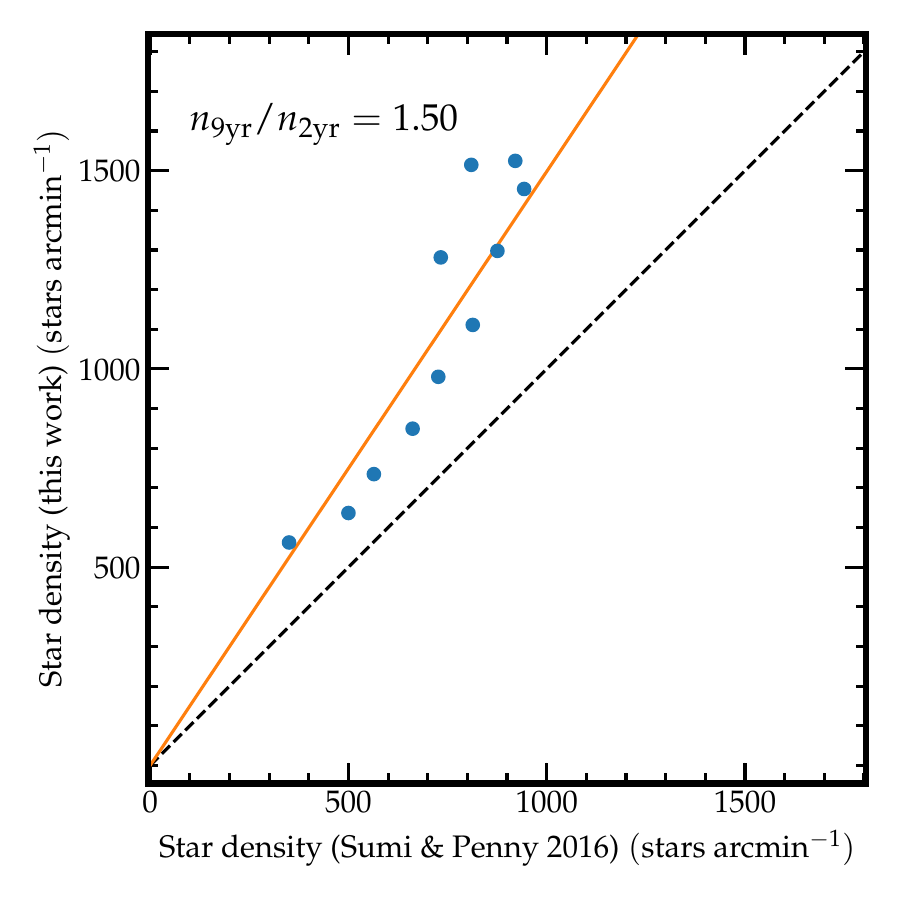}
\caption{
A comparison of the surface star density for the central region with $|l|<3^\circ$ between the dataset used in \citet{sum16} and this study.
Subfields are binned with a bin width of $\Delta b = 30\arcmin$.
\label{fig-ns_comp}}
\end{figure*}

\section{Discussion} \label{sec-dis}
\subsection{Comparison with previous results} \label{sec-comp_moa}
\citet{mro19} compared their measurement of the optical depth, $\tau$, and event rate, $\Gamma$, based on the data from 8 years of OGLE-IV observation with the result from the MOA-II 2 year survey \citep{sum16}.
As a result, they found that $\tau$ and $\Gamma$ from OGLE-IV are factors of $\sim 1.4$ lower than those based on the MOA-II all source events.
We compare our new measurements from the MOA-II 9 year data with those previous results.

Because \citet{sum16} use only events with $I_{\rm s}<20 ~\rm mag$ and $t_{\rm E} < 200 \rm ~ days$, we follow suit and re-selected only events that satisfy the same criteria in our sample of $3525$ events.
As a result of this cut, a total of $2436$ events remained in our sample.
We also re-calculate detection efficiencies using only samples with $I_{\rm s} < 20 ~\rm mag$ in the image-level simulation described in Sec. \ref{sec-eff}.

We fit the optical depth and event rate of the central region with $|l|<3^\circ$ for the reselected $2436$ sample using the same method described in Sec. \ref{sec-fit_tau} and \ref{sec-fit_gam}, and estimate that $\tau = [1.40\pm0.04]\exp[(0.36\pm0.03)(3^\circ-|b|)]\times10^{-6}$ and $\Gamma = [14.18\pm0.30]\exp[(0.42\pm0.02)(3^\circ-|b|)]\times10^{-6}$.
Figs. \ref{fig-tau_comp} and \ref{fig-gamma_comp} present the data and best-fit models from the MOA-II 9 yr sample (this work), the MOA-II 2 yr sample \citep{sum16}, and the OGLE-IV 8 yr sample \citep{mro19}.
Our measurements of $\tau$ and $\Gamma$ are systematically lower than the result from MOA-II 2 yr: $\tau = [1.84\pm0.14]\exp[(0.44\pm0.07)(3^\circ-|b|)]\times10^{-6}$ and $\Gamma = [18.74\pm0.91]\exp[(0.53\pm0.05)(3^\circ-|b|)]\times10^{-6}$ \citep{sum16} and in agreement with result from OGLE-IV 8 yr: $\tau = [1.36\pm0.04]\exp[(0.39\pm0.03)(3^\circ-|b|)]\times10^{-6}$ and $\Gamma = [13.4\pm0.3]\exp[(0.49\pm0.02)(3^\circ-|b|)]\times10^{-6}$ \citep{mro19}.

\citet{mro19} attribute the relatively high optical depth and event rate measurement from MOA-II 2yr to the incompleteness of source star counts.
Indeed, they found that their star counts were a factor of $1.5$ larger than those reported by \citet{sum16}.
Fig. \ref{fig-ns_comp} compares the surface density of the source stars used in this study with those in \citet{sum16}.
We found that our star counts are a factor of $1.5$ larger than those reported by \citet{sum16} which confirmed \citet{mro19}.
This is due to the improvements in the source star counts method described in Sec. \ref{sec-ns}.
The systematic difference in star counts is sufficient to explain the systematic excess in the measurements of $\tau$ and $\Gamma$ by \citet{sum16}.
In fact, the event rate per square degree per year, $\Gamma_{\deg^{2}}$, which is independent of the number of source stars is roughly consistent with the results of \citet{sum16}. 
Therefore, we conclude that the differences in the measurements of $\tau$ and $\Gamma$ are due to the differences in source star counts.
These systematics have been corrected in this work.

\begin{figure*}
\centering
\includegraphics[width=15cm]{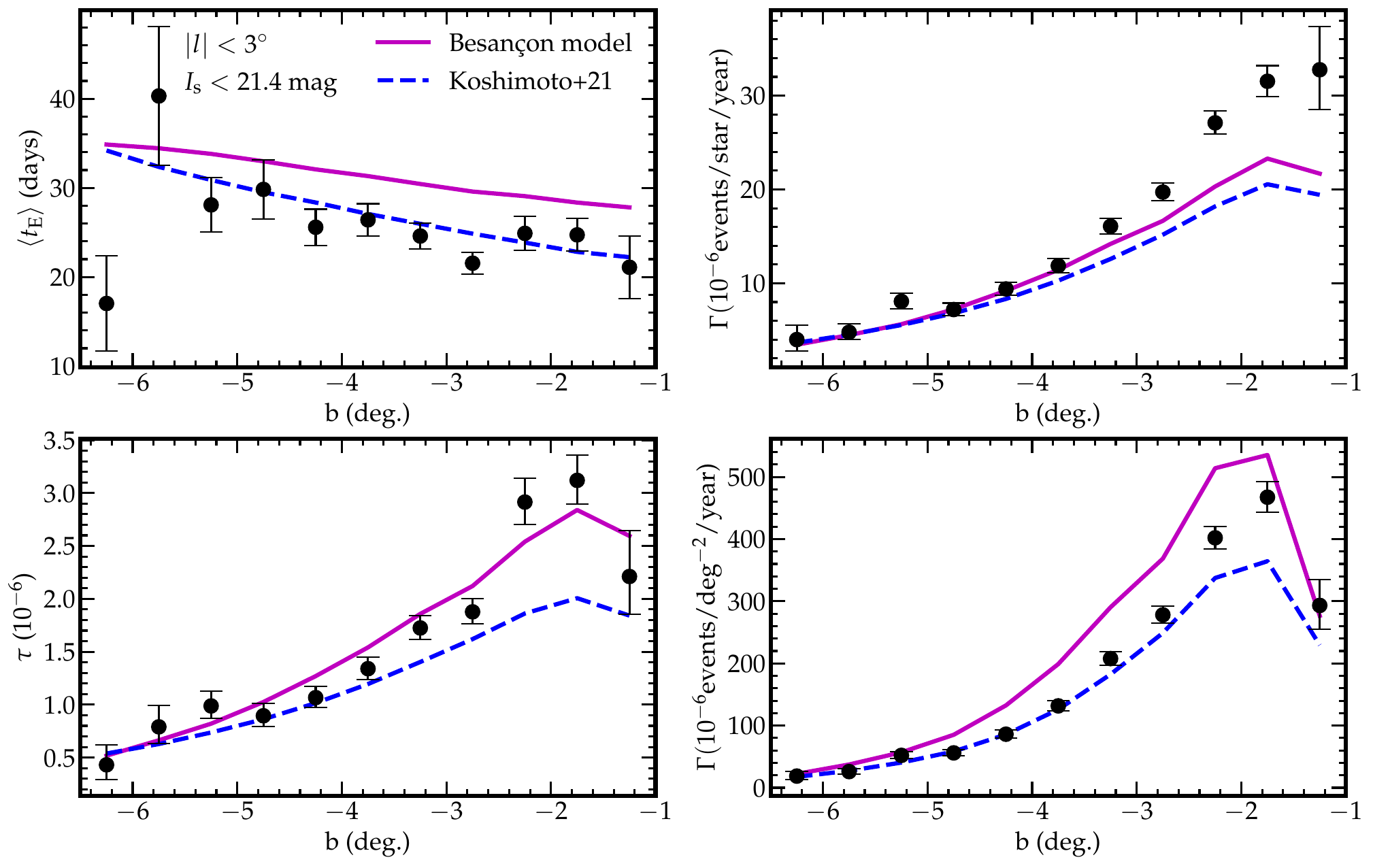}
\caption{
Comparison of the microlensing optical depth, event rate, and mean Einstein radius crossing time in the central region $(l < 3^\circ)$ with $I<21.4 ~\rm mag$ between our result and values predicted by the Galactic model.
The purple solid line shows the prediction based on MaB$\mu$lS-2 \citep{spe20} and the blue dotted line shows the prediction based on the modified version of \texttt{genulens} \citep{kos22}
\label{fig-comp_model_21.4}}
\end{figure*}

\begin{figure*}
\centering
\includegraphics[width=15cm]{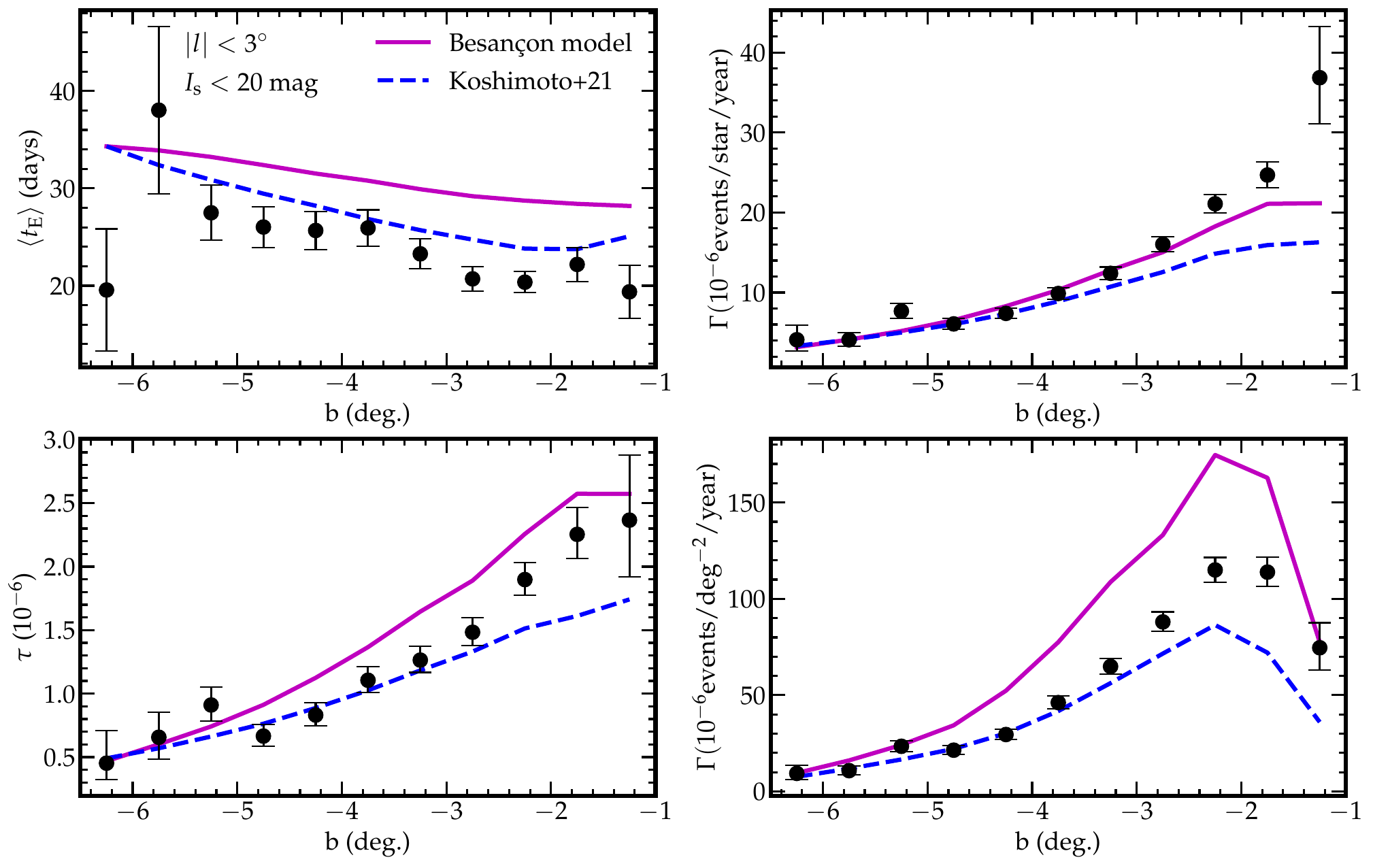}
\caption{
Comparison of the microlensing optical depth, event rate, and mean Einstein radius crossing time in the central region $(l < 3^\circ)$ with $I<20 ~\rm mag$ between our result and values predicted by the Galactic model.
The line styles follow those used in Fig. \ref{fig-comp_model_21.4}
\label{fig-comp_model_20}}
\end{figure*}

\begin{figure*}
\centering
\includegraphics[width=19cm]{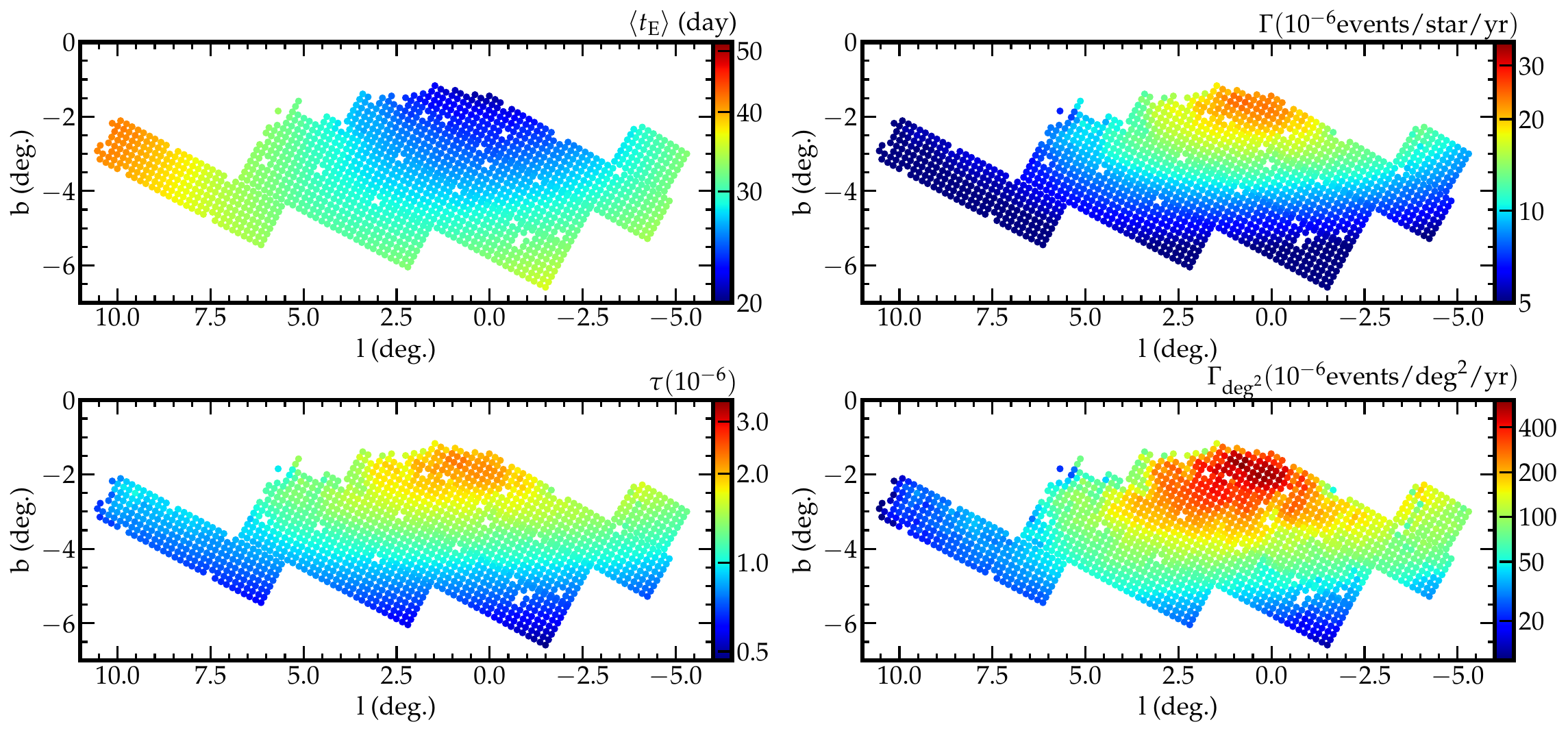}
\caption{
Color map of the mean Einstein radius crossing time, $\langle t_{\rm E} \rangle$ (top left), the microlensing optical depth, $\tau$, (bottom left), and the event rate per tar per year, $\Gamma$, (top right), the event rate per square degree per year, $\Gamma_{\deg^{2}}$, (bottom right) predicted by \texttt{genulens} \citep{kos22}.
\label{fig-kosmodel_map}}
\end{figure*}

\subsection{Comparison with Galactic Models}\label{sec-comp_galmodel}
We compare our measurement of the optical depth, event rate, and the mean Einstein radius crossing time with values predicted by models.
For this comparison, we employ two microlensing simulators: the second-generation Manchester-Besan\c{c}on Microlensing Simulator (MaB$\mu$lS-2\footnote{http://www.mabuls.net/.}) from \citet{spe20}, and a modified version of  \texttt{genulens} \citep{kos22}.

MaB$\mu$lS-2 is based on the Besan\c{c}on Galactic synthesis model \citep{rob14}.
It provides high-resolution maps of microlensing optical depth, event rate, and average timescales across a $400 ~\rm deg^2$ region of the Galactic bulge, factoring in both unresolved stellar backgrounds and limb-darkened source profiles.
The first field-by-field comparison between the prediction from the Besan\c{c}on model and observational results was conducted by \citet{awi16}.
They compared the microlensing optical depth, $\tau$, and event rate per star per year, $\Gamma$, measured by MOA-II 2 yr \citep{sum13} with values predicted by the previous version of MaB$\mu$lS and found that measured $\tau$ and $\Gamma$ at low Galactic latitude $(|l|<3^\circ)$ were about $50\%$ higher than predicted, likely due to underestimated extinction and star counts in the innermost regions.
To address this issue, \citet{sum16} corrected the incompleteness in their red clump giant counts used to normalize the luminosity function, leading to a revised estimation of $\tau$ and $\Gamma$.
Despite this correction, the revised values of $\tau$ and $\Gamma$ were still a factor of $1.5$ larger than the predictions of the Besan\c{c}on model by \citet{awi16}.
However, the newer MaB$\mu$lS-2 model, when compared to the efficiency-corrected OGLE-IV event sample, shows much better agreement, even capturing small-scale features in the event rate map \citep{spe20}.

The second model we use for comparison is a modified version of \texttt{genulens} \citep{kos22} which is a microlensing event simulation tool based on the Galactic model developed by \citet{kos21}.
This model was designed to reproduce the stellar distribution toward the Galactic bulge by fitting to the Gaia DR2 velocity data \citep{kat18}, OGLE-III red clump star count data \citep{nat13}, VIRAC proper motion catalog \citep{smi18, cla19}, BRAVA radial velocity measurements \citep{ric07, kun12}, and OGLE-IV star count and microlensing rate data \citep{mro17, mro19}.
Recently, \citet{nun24} confirmed that this model can almost perfectly reproduce the two-dimensional distribution of $t_E$ and $\mu_{\rm rel}$ from the MOA-II 9 yr FSPL dataset \citep{kos23}.

Figs. \ref{fig-comp_model_21.4} and \ref{fig-comp_model_20} compare the optical depth, event rate, and the mean Einstein radius crossing time predicted by two models with our measurement with $I < 21.4~\rm mag$ and $I < 20~\rm mag$, respectively.
Comparison of the models and data in both samples -- $I < 21.4~\rm mag$ and $I < 20~\rm mag$ -- are consistent and do not vary significantly between these two cases.

Regarding the mean Einstein radius crossing time, $\langle t_{\rm E} \rangle$, MaB$\mu$lS-2 systematically predicts higher values compared to the data, whereas \texttt{genulens} provides predictions that are in good agreement with our measurements.
The systematic overestimation of the Einstein radius crossing time in the MaB$\mu$lS-2 is similarly mentioned in \citet{spe20}.

On the other hand, neither model adequately explains the measured optical depth.
There is a systematic excess in the data compared to the predictions from \texttt{genulens} in the central region with $|b| < 3^\circ$, whereas the model performs well for $|b| > 3^\circ$.
MaB$\mu$lS-2 systematically predicts slightly higher values for $|b| > 3^\circ$ but shows better predictions compared to \texttt{genulens} for $|b| < 3^\circ$.

In contrast to the differing predictions for the mean Einstein radius crossing time and optical depth, both models predict nearly the same values for the event rate per star per year.
These predictions are consistent with our observation for $|b| > 3^\circ$, but they fail to reproduce the higher observed values in the central region with $|b| < 3^\circ$.
This systematic excess is also mentioned \citet{spe20}.
Although both models show systematic differences from the data, MaB$\mu$lS-2 provides slightly better predictions compared to \texttt{genulens} for the event rate per star per year.

Regarding the event rate per square degree per year, the two models exhibit opposing behaviors relative to the data.
While MaB$\mu$lS-2 tends to slightly overestimate the event rate, \texttt{genulens} shows a tendency to slightly underestimate it.
This tendency is more pronounced in the central region $|b| < 3^\circ$, and in fact, for $|b| > 3^\circ$, \texttt{genulens} provides good predictions.

We present color maps of predictions from \texttt{genulens} in Fig. \ref{fig-kosmodel_map}.

\section{Summary}\label{sec-conclu}
We measured the microlensing optical depth, $\tau$, and event rate, $\Gamma$, toward the Galactic bulge using the data set from the MOA-II survey from 2006 to 2014.
Our sample consists of $3525$ microlens events, with Einstein radius crossing time of $ 1~\rm day < t_{\rm E} < 760~\rm days$.
Our result is $\tau = [1.75\pm0.04]\exp[{(0.34\pm0.02)(3^{\circ}-|b|)}]\times 10^{-6}$ and $\Gamma = [16.08\pm0.28]\exp[{(0.44\pm0.02)(3^{\circ}-|b|)}]\times 10^{-6}$ in the central region with $|l|<5^\circ$. 

These results are consistent with the latest OGLE-IV 8 year data set\citep{mro19}. We confirmed that the factor of $\sim 1.4$  higher optical depth and event rate of the MOA-II 2 year result compared to the OGLE-IV 8 year result is due to a factor of $1.5$ underestimate of the source star counts in MOA analysis.

We also compared our results with model predictions, MaB$\mu$lS-2 \citep{spe20}, and a modified version of  \texttt{genulens} \citep{kos22}.
We found a systematic discrepancy between the two models and our observed values, especially in the central region with $|b|<3^\circ$.
Both models may need to be updated based on this result.

Microlensing event samples in Galactic central regions are expected to increase with future microlensing surveys.
The PRime-focus Infrared Microlensing Experiment (PRIME) began their survey toward the Galactic bulge and center in 2023 \citep{kon23,yam23}.
PRIME is expected to discover $\sim 3900$ microlensing events per year within $|b|<3^\circ$ \citep{kon23}.
In addition, the Nancy Grace Roman Space Telescope is planned to launch in late 2026 \citep{spe15} and a total of $\sim 27000$ microlensing events with $|u_{0}|<1$ are expected to be discovered \citep{pen19}.This appendix provides detailed data tables for each subfield analyzed in this study. These tables contain the key measurements and results for all subfields, supporting the main findings
It is anticipated that we can improve this work, like 2D fitting to the event rate map, by using these large samples in the future.

\begin{acknowledgments}
KN was supported by Ono Scholarship Foundation for a public interest incorporated foundation.
The work of N. Koshimoto was supported JSPS KAKENHI grant Nos. JP24K17089 and JP23KK0060.
\end{acknowledgments}

\appendix
\section{Detailed Data Tables for Each Subfield}\label{sec-subfield}
\begin{deluxetable}{cccccccccc}
    \tablewidth{10pt} 
    \tablenum{3}
    \tablecaption{MOA-II Galactic Bulge Sub Fields with Galactic Coordinates ($l$,$b$), the Number of Source Stars with $I_{\rm s}<21.4~\rm mag$  ($N_{\rm s}$), the Number of Microlensing Events ($N_{\rm ev}$), the Microlensing Event Rate per Star per Year ($\Gamma$), the Microlensing Event Rate per Square Degree per Year ($\Gamma_{\rm deg^2}$), the Optical Depth ($\tau$), the average Einstein radius crossing time $(\langle t_{\rm E} \rangle)$ average, and the average detection efficiency $(\langle \epsilon \rangle)$. \label{table-allsubfield}}
    \tablehead{
    \colhead{Field} & \colhead{$l$} & \colhead{$b$} & \colhead{$N_{\rm s}$} & \colhead{{$N_{\rm eve}$}} & \colhead{$\tau$} & \colhead{$\Gamma$} & \colhead{$\Gamma_{\rm deg^2}$} & \colhead{$\langle t_{\rm E}\rangle$}& \colhead{$\langle \epsilon \rangle$} \\
    \colhead{} & \colhead{($^\circ$)} & \colhead{($^\circ$)} & \colhead{$(10^{5}$)} & \colhead{} & \colhead{$(10^{-6}$)} & \colhead{$(10^{-6}\rm star ^{-1} year ^{-1})$} & \colhead{$(\rm deg ^{-2} year ^{-1})$} & \colhead{$(\rm days)$} & \colhead{$(10^{-2}$)}
    }
    \startdata 
    gb1-1-0 & $-4.10$ & $-2.28$ & $2.29$ & $2$ &  $2.52^{+3.81}_{-1.26}$ & $11.75^{+2.71}_{-2.36}$ & $98.73^{+22.80}_{-19.83}$ & $48.46{\pm18.74}$ & $8.61$\\
    gb1-1-1 & $-4.02$ & $-2.42$ & $3.30$ & $4$ &  $2.32^{+1.67}_{-1.16}$ & $14.35^{+2.44}_{-2.20}$ & $173.91^{+29.60}_{-26.62}$ & $36.49{\pm2.22}$ & $8.59$\\
    gb1-1-2 & $-3.93$ & $-2.56$ & $1.67$ & $0$ & $\cdots$ & $\cdots$ & $\cdots$ & $\cdots$ & $8.57$\\
    gb1-1-3 & $-3.84$ & $-2.71$ & $2.15$ & $1$ &  $1.01^{+2.10}_{-0.07}$ & $5.24^{+1.97}_{-1.60}$ & $41.22^{+15.54}_{-12.57}$ & $43.70$ & $8.56$\\
    gb1-1-5 & $-3.87$ & $-2.34$ & $2.01$ & $1$ &  $1.14^{+2.39}_{-0.11}$ & $5.51^{+2.09}_{-1.69}$ & $40.71^{+15.46}_{-12.49}$ & $46.93$ & $8.55$\\
    gb1-1-6 & $-3.79$ & $-2.48$ & $1.38$ & $0$ & $\cdots$ & $\cdots$ & $\cdots$ & $\cdots$ & $8.53$\\
    gb1-1-7 & $-3.70$ & $-2.62$ & $2.61$ & $2$ &  $2.02^{+3.01}_{-0.97}$ & $8.49^{+2.18}_{-1.87}$ & $81.20^{+20.88}_{-17.91}$ & $53.70{\pm8.42}$ & $8.52$\\
    gb1-2-0 & $-4.40$ & $-2.46$ & $3.04$ & $3$ &  $1.66^{+1.71}_{-0.60}$ & $17.08^{+2.77}_{-2.50}$ & $190.37^{+30.88}_{-27.90}$ & $21.92{\pm9.26}$ & $8.69$\\
    gb1-2-1 & $-4.32$ & $-2.60$ & $3.42$ & $3$ &  $1.03^{+0.97}_{-0.37}$ & $14.28^{+2.39}_{-2.16}$ & $179.04^{+30.00}_{-27.03}$ & $16.30{\pm3.15}$ & $8.67$\\
    gb1-2-2 & $-4.23$ & $-2.74$ & $3.49$ & $3$ &  $1.34^{+1.40}_{-0.47}$ & $14.05^{+2.35}_{-2.12}$ & $179.72^{+30.06}_{-27.08}$ & $21.62{\pm8.14}$ & $8.65$\\
    gb1-2-3 & $-4.14$ & $-2.89$ & $3.18$ & $4$ &  $3.60^{+2.66}_{-1.73}$ & $14.78^{+2.53}_{-2.27}$ & $172.28^{+29.47}_{-26.49}$ & $55.07{\pm12.49}$ & $8.64$\\
    gb1-2-4 & $-4.26$ & $-2.37$ & $3.22$ & $3$ &  $1.58^{+1.57}_{-0.56}$ & $12.36^{+2.31}_{-2.06}$ & $146.04^{+27.29}_{-24.31}$ & $28.91{\pm6.31}$ & $8.65$\\
    gb1-2-5 & $-4.17$ & $-2.51$ & $3.47$ & $5$ &  $1.79^{+1.09}_{-0.72}$ & $20.00^{+2.78}_{-2.55}$ & $254.51^{+35.40}_{-32.43}$ & $20.29{\pm2.42}$ & $8.63$\\
    gb1-2-6 & $-4.08$ & $-2.66$ & $2.53$ & $0$ & $\cdots$ & $\cdots$ & $\cdots$ & $\cdots$ & $8.61$\\
    gb1-2-7 & $-4.00$ & $-2.80$ & $0.85$ & $1$ &  $1.10^{+2.35}_{-0.08}$ & $17.31^{+5.60}_{-4.65}$ & $53.75^{+17.41}_{-14.44}$ & $14.38$ & $8.60$\\
    gb1-3-0 & $-4.70$ & $-2.64$ & $3.19$ & $4$ &  $2.76^{+1.95}_{-1.37}$ & $15.19^{+2.56}_{-2.30}$ & $177.64^{+29.89}_{-26.92}$ & $41.07{\pm6.37}$ & $8.76$\\
    gb1-3-1 & $-4.62$ & $-2.78$ & $3.21$ & $6$ &  $4.89^{+3.24}_{-1.68}$ & $22.22^{+3.05}_{-2.79}$ & $261.47^{+35.86}_{-32.88}$ & $49.77{\pm9.44}$ & $8.74$\\
    gb1-3-2 & $-4.53$ & $-2.92$ & $3.08$ & $5$ &  $2.45^{+1.86}_{-1.01}$ & $23.79^{+3.22}_{-2.95}$ & $268.57^{+36.32}_{-33.34}$ & $23.28{\pm5.38}$ & $8.73$\\
    gb1-3-3 & $-4.44$ & $-3.07$ & $3.09$ & $2$ &  $0.28^{+0.28}_{-0.14}$ & $20.75^{+3.01}_{-2.75}$ & $235.44^{+34.12}_{-31.15}$ & $3.09{\pm0.07}$ & $8.72$\\
    gb1-3-4 & $-4.56$ & $-2.55$ & $2.96$ & $3$ &  $3.37^{+3.30}_{-1.22}$ & $11.41^{+2.33}_{-2.05}$ & $124.05^{+25.31}_{-22.34}$ & $66.71{\pm13.60}$ & $8.73$\\
    gb1-3-5 & $-4.47$ & $-2.69$ & $3.24$ & $2$ &  $0.90^{+0.92}_{-0.45}$ & $7.94^{+1.88}_{-1.63}$ & $94.43^{+22.35}_{-19.37}$ & $25.72{\pm1.89}$ & $8.71$\\
    gb1-3-6 & $-4.39$ & $-2.84$ & $3.33$ & $6$ &  $2.68^{+1.36}_{-0.87}$ & $24.26^{+3.11}_{-2.87}$ & $296.75^{+38.08}_{-35.10}$ & $24.94{\pm3.30}$ & $8.69$\\
    gb1-3-7 & $-4.30$ & $-2.98$ & $3.13$ & $2$ &  $2.06^{+2.09}_{-0.99}$ & $6.51^{+1.75}_{-1.49}$ & $74.81^{+20.13}_{-17.16}$ & $71.58{\pm1.68}$ & $8.68$\\
    \enddata
    \tablecomments{This table is available in a machine-readable form.}
    \end{deluxetable}

This appendix provides detailed data tables for each subfield analyzed in this study. 
Table \ref{table-allsubfield} contains the key measurements and results for all subfields, supporting the main findings.


\begin{thebibliography}{}
    \bibitem[Afonso et al.(2003)]{afa03} Afonso, C., Albert, J.~N., Alard, C., et al.\ 2003, \aap, 404, 145. doi:10.1051/0004-6361:20030307
    \bibitem[Alard(2000)]{ala00} Alard, C.\ 2000, \aaps, 144, 363. doi:10.1051/aas:2000214
    \bibitem[Alard \& Lupton(1998)]{ala96} Alard, C. \& Lupton, R.~H.\ 1998, \apj, 503, 325. doi:10.1086/305984
    \bibitem[Alcock et al.(1997)]{alc97} Alcock, C., Allsman, R.~A., Alves, D., et al.\ 1997, \apj, 486, 697. doi:10.1086/304535
    \bibitem[Alcock et al.(2000)]{alc00} Alcock, C., Allsman, R.~A., Alves, D.~R., et al.\ 2000, \apj, 541, 734. doi:10.1086/309484
    \bibitem[Awiphan et al.(2016)]{awi16} Awiphan, S., Kerins, E., \& Robin, A.~C.\ 2016, \mnras, 456, 1666. doi:10.1093/mnras/stv2625
    \bibitem[Batista et al.(2011)]{bat11} Batista, V., Gould, A., Dieters, S., et al.\ 2011, \aap, 529, A102. doi:10.1051/0004-6361/201016111
    \bibitem[Bond et al.(2001)]{bon01} Bond, I.~A., Abe, F., Dodd, R.~J., et al.\ 2001, \mnras, 327, 868. doi:10.1046/j.1365-8711.2001.04776.x
    \bibitem[Clarke et al.(2019)]{cla19} Clarke, J.~P., Wegg, C., Gerhard, O., et al.\ 2019, \mnras, 489, 3519. doi:10.1093/mnras/stz2382
    \bibitem[Griest et al.(1991)]{gri91} Griest, K., Alcock, C., Axelrod, T.~S., et al.\ 1991, \apjl, 372, L79. doi:10.1086/186028
    \bibitem[Gyuk(1999)]{gyu99} Gyuk, G.\ 1999, \apj, 510, 205. doi:10.1086/306544
    \bibitem[Gaia Collaboration et al.(2018)]{kat18} Gaia Collaboration, Katz, D., Antoja, T., et al.\ 2018, \aap, 616, A11
    \bibitem[Hamadache et al.(2006)]{ham06} Hamadache, C., Le Guillou, L., Tisserand, P., et al.\ 2006, \aap, 454, 185. doi:10.1051/0004-6361:20064893
    \bibitem[Holtzman et al.(1998)]{hol98} Holtzman, J.~A., Watson, A.~M., Baum, W.~A., et al.\ 1998, \aj, 115, 1946. doi:10.1086/300336
    \bibitem[Kim et al.(2016)]{kim16} Kim, S.-L., Lee, C.-U., Park, B.-G., et al.\ 2016, Journal of Korean Astronomical Society, 49, 37. doi:10.5303/JKAS.2016.49.1.37
    \bibitem[Kiraga et al.(1997)]{kir97} Kiraga, M., Paczy{\'n}ski, B., \& Stanek, K.~Z.\ 1997, \apj, 485, 611. doi:10.1086/304441
    \bibitem[Koshimoto et al.(2021)]{kos21} Koshimoto, N., Baba, J., \& Bennett, D.~P.\ 2021b, \apj, 917, 78. doi:10.3847/1538-4357/ac07a8 
    \bibitem[Koshimoto \& Ranc(2022)]{kos22} Koshimoto, N. \& Ranc, C.\ 2022, Zenodo.4784948
    \bibitem[Koshimoto et al.(2023)]{kos23} Koshimoto, N., Sumi, T., Bennett, D.~P., et al.\ 2023, \aj, 166, 107. doi:10.3847/1538-3881/ace689
    \bibitem[Kiraga \& Paczynski(1994)]{kir94} Kiraga, M. \& Paczynski, B.\ 1994, \apjl, 430, L101. doi:10.1086/187448
    \bibitem[Kondo et al.(2023)]{kon23} Kondo, I., Sumi, T., Koshimoto, N., et al.\ 2023, \aj, 165, 254. doi:10.3847/1538-3881/acccf9
    \bibitem[Kunder et al.(2012)]{kun12} Kunder, A., Koch, A., Rich, R.~M., et al.\ 2012, \aj, 143, 57. doi:10.1088/0004-6256/143/3/57
    \bibitem[Mr{\'o}z et al.(2017)]{mro17} Mr{\'o}z, P., Udalski, A., Skowron, J., et al.\ 2017, \nat, 548, 183 
    \bibitem[Mr{\'o}z et al.(2019)]{mro19} Mr{\'o}z, P., Udalski, A., Skowron, J., et al.\ 2019, \apjs, 244, 29. doi:10.3847/1538-4365/ab426b
    \bibitem[Nataf et al.(2013)]{nat13} Nataf, D.~M., Gould, A., Fouqu{\'e}, P., et al.\ 2013, \apj, 769, 88. doi:10.1088/0004-637X/769/2/88
    \bibitem[Nunota et al.(2024)]{nun24} Nunota, K., Koshimoto, N., Suzuki, D., et al.\ 2024, \apj, 967, 77. doi:10.3847/1538-4357/ad3cdc
    \bibitem[Paczynski(1986)]{pac86} Paczynski, B.\ 1986, \apj, 304, 1. doi:10.1086/164140
    \bibitem[Paczynski(1991)]{pac91} Paczynski, B.\ 1991, \apjl, 371, L63. doi:10.1086/186003
    \bibitem[Paczynski et al.(1994)]{pac94} Paczynski, B., Stanek, K.~Z., Udalski, A., et al.\ 1994, \apjl, 435, L113. doi:10.1086/187607
    \bibitem[Peale(1998)]{pea98} Peale, S.~J.\ 1998, \apj, 509, 177. doi:10.1086/306490
    \bibitem[Penny et al.(2019)]{pen19} Penny, M.~T., Gaudi, B.~S., Kerins, E., et al.\ 2019, \apjs, 241, 3. doi:10.3847/1538-4365/aafb69
    \bibitem[Popowski et al.(2001)]{pop01} Popowski, P., Alcock, C., Allsman, R.~A., et al.\ 2001, Microlensing 2000: A New Era of Microlensing Astrophysics, 239, 244. doi:10.48550/arXiv.astro-ph/0005466
    \bibitem[Popowski et al.(2005)]{pop05} Popowski, P., Griest, K., Thomas, C.~L., et al.\ 2005, \apj, 631, 879. doi:10.1086/432246
    \bibitem[Sako et al.(2008)]{sak08} Sako, T., Sekiguchi, T., Sasaki, M., et al.\ 2008, Experimental Astronomy, 22, 51. doi:10.1007/s10686-007-9082-5
    \bibitem[Robin et al.(2014)]{rob14} Robin, A.~C., Reyl{\'e}, C., Fliri, J., et al.\ 2014, \aap, 569, A13. doi:10.1051/0004-6361/201423415
    \bibitem[Rich et al.(2007)]{ric07} Rich, R.~M., Reitzel, D.~B., Howard, C.~D., et al.\ 2007, \apjl, 658, L29. doi:10.1086/513509
    \bibitem[Smith et al.(2018)]{smi18} Smith, L.~C., Lucas, P.~W., Kurtev, R., et al.\ 2018, \mnras, 474, 1826. doi:10.1093/mnras/stx2789
    \bibitem[Specht et al.(2020)]{spe20} Specht, D., Kerins, E., Awiphan, S., et al.\ 2020, \mnras, 498, 2196. doi:10.1093/mnras/staa2375
    \bibitem[Spergel et al.(2015)]{spe15} Spergel, D., Gehrels, N., Baltay, C., et al.\ 2015, arXiv:1503.03757. doi:10.48550/arXiv.1503.03757
    \bibitem[Stanek et al.(2000)]{sta00} Stanek, K.~Z., Kaluzny, J., Wysocka, A., et al.\ 2000, \actaa, 50, 191. doi:10.48550/arXiv.astro-ph/9908041
    \bibitem[Stetson(1987)]{ste87} Stetson, P.~B.\ 1987, \pasp, 99, 191. doi:10.1086/131977
    \bibitem[Sumi et al.(2003)]{sum03} Sumi, T., Abe, F., Bond, I.~A., et al.\ 2003, \apj, 591, 204. doi:10.1086/375212
    \bibitem[Sumi et al.(2013)]{sum13} Sumi, T., Bennett, D.~P., Bond, I.~A., et al.\ 2013, \apj, 778, 150. doi:10.1088/0004-637X/778/2/150
    \bibitem[Sumi \& Penny(2016)]{sum16} Sumi, T. \& Penny, M.~T.\ 2016, \apj, 827, 139. doi:10.3847/0004-637X/827/2/139
    \bibitem[Sumi et al.(2023)]{sum23} Sumi, T., Koshimoto, N., Bennett, D.~P., et al.\ 2023, \aj, 166, 108. doi:10.3847/1538-3881/ace688
    \bibitem[Szyma{\'n}ski et al.(2011)]{szy11} Szyma{\'n}ski, M.~K., Udalski, A., Soszy{\'n}ski, I., et al.\ 2011, \actaa, 61, 83. doi:10.48550/arXiv.1107.4008
    \bibitem[Tomaney \& Crotts(1996)]{tom96} Tomaney, A.~B. \& Crotts, A.~P.~S.\ 1996, \aj, 112, 2872. doi:10.1086/118228
    \bibitem[Udalski et al.(1994)]{uda94} Udalski, A., Szymanski, M., Stanek, K.~Z., et al.\ 1994, \actaa, 44, 165. doi:10.48550/arXiv.astro-ph/9407014
    \bibitem[Yama et al.(2023)]{yam23} Yama, H., Suzuki, D., Miyazaki, S., et al.\ 2023, Journal of Astronomical Instrumentation, 12, 2350004. doi:10.1142/S2251171723500046
    \bibitem[Zhao et al.(1995)]{zha95} Zhao, H., Spergel, D.~N., \& Rich, R.~M.\ 1995, \apjl, 440, L13. doi:10.1086/187749
    \bibitem[Zhao \& Mao(1996)]{zha96} Zhao, H. \& Mao, S.\ 1996, \mnras, 283, 1197. doi:10.1093/mnras/283.4.1197

    \end{thebibliography}
\end{document}